\documentclass[reprint,preprintnumbers,amsmath,amssymb,superscriptaddress]{revtex4-1}
\usepackage{graphicx}% Include figure files
\usepackage{dcolumn}% Align table columns on decimal point
\usepackage{bm}% bold math
\usepackage{float}
\usepackage{natbib}
\usepackage{color}

\begin{document}
%\captionsetup{justification=raggedright, singlelinecheck=false,belowskip=-13pt,aboveskip=4pt}

\title{Spin Seebeck imaging of spin-torque switching in antiferromagnetic Pt/NiO heterostructures}% Force line breaks with \\

\author{Isaiah Gray}
\affiliation{School of Applied and Engineering Physics, Cornell University, Ithaca NY 14853}
\affiliation{Kavli Institute for Nanoscale Science, Ithaca NY 14853}
\author{Takahiro Moriyama}
\affiliation{Institute for Chemical Research, Kyoto University, Uji, Kyoto, 611-0011, Japan}
\author{Nikhil Sivadas}
\affiliation{School of Applied and Engineering Physics, Cornell University, Ithaca NY 14853}
\author{Gregory M. Stiehl}
\affiliation{Department of Physics, Cornell University, Ithaca NY 14853}
\author{John T. Heron}
\affiliation{Department of Materials Science and Engineering, University of Michigan, Ann Arbor, Michigan}
\author{Ryan Need}
\affiliation{NIST Center for Neutron Research, National Institute of Standards and Technology, Gaithersburg, Maryland 20899}
\author{Brian J. Kirby}
\affiliation{NIST Center for Neutron Research, National Institute of Standards and Technology, Gaithersburg, Maryland 20899}
\author{David H. Low}
\affiliation{School of Applied and Engineering Physics, Cornell University, Ithaca NY 14853}
\author{Katja C. Nowack}
\affiliation{Department of Physics, Cornell University, Ithaca NY 14853}
\affiliation{Kavli Institute for Nanoscale Science, Ithaca NY 14853}
\author{Darrell G. Schlom}
\affiliation{Department of Materials Science and Engineering, Cornell University, Ithaca, NY 14853}
\affiliation{Kavli Institute for Nanoscale Science, Ithaca NY 14853}
\author{Daniel C. Ralph}
\affiliation{Department of Physics, Cornell University, Ithaca NY 14853}
\affiliation{Kavli Institute for Nanoscale Science, Ithaca NY 14853}
\author{Teruo Ono}
\affiliation{Institute for Chemical Research, Kyoto University, Uji, Kyoto, 611-0011, Japan}
\author{Gregory D. Fuchs}
\affiliation{School of Applied and Engineering Physics, Cornell University, Ithaca NY 14853}
\affiliation{Kavli Institute for Nanoscale Science, Ithaca NY 14853}

\date{\today}
\begin{abstract}

As electrical control of N\'eel order opens the door to reliable antiferromagnetic spintronic devices, understanding the microscopic mechanisms of antiferromagnetic switching is crucial. Spatially-resolved studies are necessary to distinguish multiple nonuniform switching mechanisms; however, progress has been hindered by the lack of tabletop techniques to image the N\'eel order. We demonstrate spin Seebeck microscopy as a sensitive, table-top method for imaging antiferromagnetic order in thin films, and apply this technique to study spin-torque switching in NiO/Pt and Pt/NiO/Pt heterostructures. We establish the interfacial antiferromagnetic spin Seebeck effect in NiO as a probe of surface N\'eel order, resolving antiferromagnetic spin domains within crystalline twin domains. By imaging before and after applying current-induced spin torque, we resolve spin domain rotation and domain wall motion, acting simultaneously. We correlate the changes in spin Seebeck images with electrical measurements of the average N\'eel orientation through the spin Hall magnetoresistance, confirming that we image antiferromagnetic order.
\end{abstract}

\maketitle 

\section{Introduction} 
 
Antiferromagnets (AFs), long relegated to a supporting role as the pinning layers in ferromagnetic spintronic devices \cite{DienyPRB, FukeJAP}, are emerging as the active element in antiferromagnetic spintronic devices \cite{BaltzRevModPhys, JungwirthNatNano, ZeleznyNatPhys}. In constrast to ferromagnets (FMs), AFs are insensitive to magnetic fields \cite{BogdanovPRB} and exhibit dynamics at the THz frequency scale \cite{OlejnikSciAdv, BowlanJPhysD}.  Additionally, AFs have magnetotransport effects that enable electrical readout \cite{FinaNatComm, MartiNatMater}. Taking advantage of these attractive properties, however, requires overcoming the challenge of reliably manipulating the N\'eel order. 

Recent breakthroughs in electrical \cite{WadleyScience, OlejnikNatComm} and optical \cite{HiguchiNatComm} control provide a path toward reliable devices. In particular, electrical switching was demonstrated in the metals CuMnAs \cite{WadleyScience, WadleyNatNano} and $\mathrm{Mn}_2 \mathrm{Au}$ \cite{BodnarNatComm} using N\'eel spin-orbit torque, in which the sign of the spin-orbit field from DC current within the material alternates on each lattice site to coherently rotate the N\'eel vector \cite{ZeleznyPRL}. Recently, electrical switching of an AF via spin-torque was also demonstrated in insulating NiO \cite{MoriyamaArxiv, ChenPRL, BaldratiArxiv} after several predictions \cite{ChengPRL, HaneyPRL}. In this mechanism, the DC current passing through an adjacent Pt layer generates a spin current through the spin Hall effect, which then exerts an antidamping torque on the spins at the Pt/NiO interface. Switching by antidamping spin torque does not require that the spin sublattices form inversion partners, which is required for N\'eel spin-orbit torque, and hence it is a more general approach that could enable all-electrical control over a wider variety of AFs. 

Previous experiments have shown that AF switching is nonuniform \cite{GrzybowskiPRL, SapozhnikPRB, WadleyNatNano} and heavily influenced by local magneto-elastic coupling \cite{GomonayJPhysCondMat}. Nominally identical samples display switching efficiency that varies by up to a factor of 7 at the same current density \cite{MoriyamaArxiv}, demonstrating a need for better understanding the switching process at the domain level. Systematic spatially-resolved studies are necessary to firmly establish the spin rotation mechanisms, the fraction of the domains that switch, and the reproducibility of switching. 

A primary challenge when imaging antiferromagnetism is to find an experimental probe that is sensitive to the N\'eel order and also provides the sub-$\mu$m resolution necessary to resolve domains. XMLD-PEEM has been the most reliable technique \cite{GrzybowskiPRL, HillebrechtPRL}; however, it requires a coherent x-ray source that is available at only a few facilities. Second-harmonic \cite{NyvltPRB, ChauleauNatMater} and quadratic magneto-optical techniques \cite{HigoNatPhoton} are available in a table-top format, but the small signal sizes create a need for background subtraction, which can be a problem because antiferromagnets are difficult to fully saturate. As an alternative, recent demonstrations of the AF anomalous Nernst effect \cite{IkhlasNatPhys} and AF spin Seebeck effect \cite{WuPRL, SekiPRL} open up the possibility of using spin-thermal effects as an imaging probe, because they can be directly sensitive to the N\'eel order \cite{BaltzRevModPhys}. Previous work both from our group and others have demonstrated high-sensitivity imaging of ferromagnetic order via the anomalous Nernst and longitudinal spin Seebeck effects \cite{TRANENatComm, TRANEPRAppl,TRANEPRB, TRANEPRAppl2, WeilerPRL}, suggesting that a practical and sensitive magneto-thermal microscope for N\'eel order can also be developed.  

\begin{figure*}[!htb]
\centering
\includegraphics[scale=0.11]{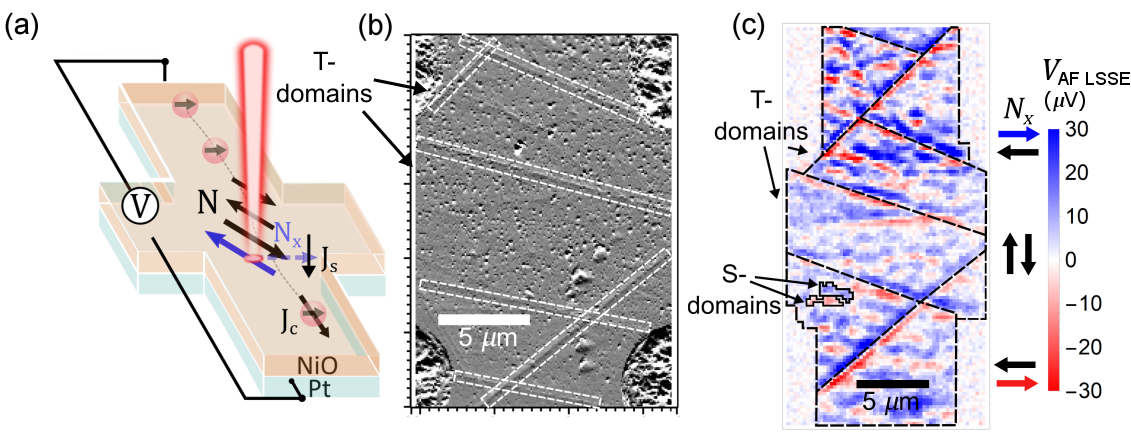}
\caption{Demonstration of antiferromagnetic longitudinal spin Seebeck effect (LSSE) microscopy. (a) Schematic of the measurement. A Ti:Sapphire laser focused to 650 nm spot size thermally generates a local spin current $\bm{J}_s$ at the Pt/NiO interface, with spin polarization $\bm{\sigma}$ parallel to the local N\'eel orientation $\bm{N}$. The sign of $\bm{\sigma}$ is determined by the spin direction of the uncompensated monolayer. $\bm{J}_s$ is transduced into a charge current via the inverse spin Hall effect in the Pt, resulting in a voltage between the contacts. We measure the projection of $\bm{N}$ transverse to the voltage contacts, which is $N_x$ in this configuration. (b) AFM amplitude image of a $10 \ \mu\mathrm{m} \times 50 \ \mu\mathrm{m}$ Hall cross of MgO/4 nm Pt/6 nm NiO(111). Diagonal lines, highlighted with white lines, indicate crystal grain (T domain) boundaries. (c) AF LSSE image of the same sample. T-domain walls are visible in AF LSSE images due to the ordinary Seebeck effect. Additional contrast within the T-domains are spin (S)-domains at the Pt/NiO: blue~(red) contrast represents spins pointing right~(left).}  
\end{figure*}

In this work we use antiferromagnetic longitudinal spin Seebeck effect (AF LSSE) microscopy to image spin-orbit torque switching in Pt/NiO(111)/Pt trilayers and NiO(111)/Pt bilayers. We provide the first experimental demonstration of interfacial AF LSSE and use it as a direct probe of the N\'eel order to resolve 1-10 $\mu$m-size  S(spin)-domains within crystalline T(twin)-domains. By imaging before and after spin-torque switching, we reveal the effects of anti-damping spin torque on the N\'eel order of NiO, showing switching by both domain rotation and domain wall motion. In particular, we show that switching occurs by continuous rotation of the N\'eel orientation rather than by flopping between in-plane easy axes, and we resolve current-polarity-dependent domain wall motion that cannot be observed with device-level transport measurements. 

The organization of this paper is as follows: we discuss the antiferromagnetic domain structure in NiO and present initial SSE images. We then establish the interfacial AF LSSE as the source of signal, and we study spin-torque-induced domain rotation and domain wall motion.

\section{Imaging N\'eel order with spin Seebeck microscopy}

\subsection{Resolving spin and twin domains in N\lowercase{i}O}

NiO is a collinear insulating antiferromagnet \cite{ChatterjiPRB} with a N\'eel temperature $T_N$ of 523 K in bulk \cite{LewisJPhysC}. Superexchange between Ni atoms along the $\langle 100 \rangle$ directions aligns the spins in ferromagnetic $\lbrace 111 \rbrace$ planes, in which spins on one plane are antiparallel to spins on the adjacent plane \cite{HillebrechtPRL}. Magnetostriction along $\langle 111 \rangle$ from the AF ordering causes crystallographic twinning, forming four T(twin)-domains in bulk NiO \cite{SlackJApplPhys} and two T-domains in $(111)$ epitaxial thin films \cite{LindahlJCrystGrowth}. Within each T-domain, dipolar next-nearest-neighbor coupling introduces a weak additional in-plane anisotropy along the three equivalent $\left[11\bar{2} \right]$ directions \cite{UchidaJPhysJapan}. In bulk crystals the spins form S(spin)-domains along these three directions \cite{SaitoJPhysC}. In thin films, however, magnetoelastic stresses, the AF equivalent of the demagnetization field in FMs \cite{GomonayJPhysCondMat}, introduce an additional effective anisotropy. This spatially inhomogeneous anisotropy pulls the spins out of well-defined $\left[11\bar{2} \right]$ directions, resulting in a disordered in-plane S-domain structure \cite{StohrPRL}. 
 
We resolve the spin domains in Pt/NiO bilayers and Pt/NiO/Pt trilayers with spin Seebeck effect microscopy \cite{TRANEPRAppl2} using a geometry illustrated in Fig. 1(a). We focus 3-ps-pulses from a Ti:Sapphire laser down to a 650 nm spot size, which produces a local out-of-plane thermal gradient, especially at the Pt/NiO interfaces (see the supporting information for more details \footnote{ Supporting information is available online at URL\label{SIref}}). We denote the interfacial temperature drop by $\Delta T$. The thermal gradient generates a local spin current $\bm{J}_s$ that diffuses into the Pt parallel to the surface normal $\hat{\mathbf{n}}$ with polarization $\bm{\sigma}$. Within the Pt, the spin current is transduced into a charge current  $\bm{J}_c \propto  \bm{J}_s \times \bm{\sigma}$ via the inverse spin Hall effect, which results in a voltage drop across the sample. By raster scanning the focused laser over the sample, we build a map of the ISHE voltage. 

\begin{figure*}[htb]
\centering
\includegraphics[scale=0.45]{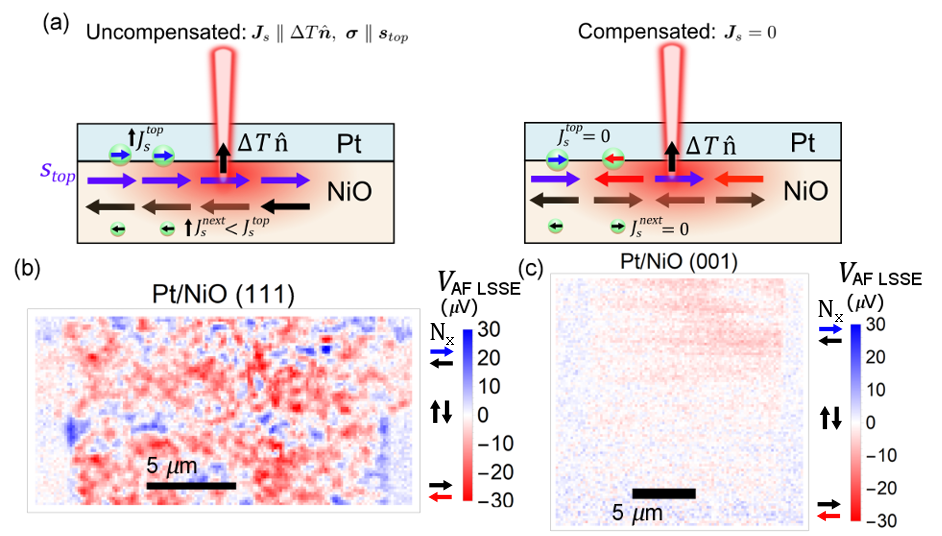}
\caption{Mechanism of the interfacial antiferromagnetic spin Seebeck effect (AF LSSE). (a) Schematic illustrating the interfacial AF LSSE. A thermal gradient localized at an uncompensated interface produces spin current with polarization $\bm{\sigma}$ parallel to the interfacial uncompensated monolayer ($\bm{s}_{top}$ in the diagram), while no spin current occurs at a compensated interface. (b,c) AF LSSE images of (c) uncompensated $\mathrm{MgAl}_2\mathrm{O}_4$(111)/165 nm NiO(111)/6 nm Pt and (d) compensated MgO(001)/136 nm NiO(001)/6 nm Pt. The lack of a signal from NiO(001) indicates that the contrast originates from the uncompensated interface. }  
\end{figure*}

An example AF LSSE image of an epitaxial MgO/4 nm Pt/6 nm NiO(111) device, deposited by sputtering and patterned into a 10~$\mu$m~$\times$~$50$~$\mu$m Hall cross by optical photolithography, is shown in Fig.~1(c), alongside the corresponding atomic force microscopy (AFM) image in Fig.~1(b). We acquire all images at room temperature and zero magnetic field using 3.4 $\mathrm{mJ}/\mathrm{cm}^2$ laser fluence. We show in the supporting information that AF LSSE signal is unaffected by magnetic field up to $\pm 250$ mT, the largest field we can apply in our setup. Note the voltage we plot is not the actual spin Seebeck voltage but rather a lock-in voltage after amplification, mixing, and normalization to account for impedance matching, which is also described in the supporting information. Characteristic sharp straight lines in the atomic force microscopy image show T-domain walls between twinned in-plane crystal grains. They are also visible in the AF LSSE images, since thermal discontinuity causes an artifact in the signal from the ordinary Seebeck effect \footnote{The laser, in addition to the vertical thermal gradient that produces the spin Seebeck effect, also generates an in-plane thermal gradient that produces electric fields from the ordinary Seebeck effect. In a thermally isotropic region, these electric fields cancel out in detail. At the T-domain wall, there is an in-plane thermal discontinuity that results in positive voltage on one side of the wall and negative voltage on the other.}. 
	
Contrast within T-domains in the AF LSSE images show S-domains: examples are highlighted in black enclosures in Fig.~1(c). The image contrast represents the in-plane component of the N\'eel vector at the NiO/Pt interface transverse to the voltage contacts, which is $N_x$ throughout this work (in the next section, we present evidence for this assertion, based on an interfacial AF LSSE as the signal source). Since we currently cannot saturate the N\'eel vector along a given direction while imaging, which would require applying an \textit{in situ} magnetic field greater than the spin-flop field (5 T in NiO \cite{MachadoPRB}), we cannot calibrate the AF LSSE voltage to an absolute N\'eel orientation. This is an intrinsic difficulty of detecting antiferromagnetism and is also experienced by other imaging techniques, including XMLD-PEEM. Instead, the AF LSSE voltage represents the strength of the projection without absolute calibration, where positive (blue) signal shows spins pointing right and negative (red) signal shows spins pointing left. The size, shape, and distribution of S- and T-domains are consistent with previous XMLD-PEEM imaging studies of thin-film NiO \cite{StohrPRL, AraiPRB}. 

\subsection{Evidence for interfacial antiferromagnetic LSSE}

We attribute the signal in our images to an interfacial AF LSSE. Although the ferromagnetic LSSE is well-established both in bulk \cite{UchidaNature, JaworskiNatMater, KehlbergerPRL} and at the interface \cite{XiaoPRB, KimlingPRL, GilesPRB}, the AF LSSE was initially predicted not to exist for a collinear AF\cite{OhnumaPRB} and was only recently observed \cite{WuPRL, SekiPRL}. In a collinear AF, the two spin sublattices produce two degenerate magnon modes \cite{SieversPhysRev}, which produce spin current in opposite directions under a thermal gradient. Therefore, unless the degeneracy is lifted there is no net spin current \cite{OhnumaPRB}. The degeneracy can be lifted in the AF bulk by applying a large magnetic field \cite{WuPRL, SekiPRL} or by exploiting anisotropies that result in additional magnon modes \cite{HolandaAPL, RezendePRB}. The degeneracy can also be lifted by inversion symmetry breaking at the interface, resulting in an interfacial AF LSSE that has been predicted \cite{BenderPRL} but not previously been reported. 

The mechanism of the interfacial AF LSSE is schematically illustrated in Fig.~2(a). AF interfaces can be uncompensated, meaning the layer closest to the surface contains an excess of one sublattice, or compensated, in which adjacent spins are antiparallel in each growth plane. Although both types of AFs have no bulk moment, an interfacial thermal gradient at the uncompensated interface will couple more strongly to the spins on the interfacial layer than to the spins on the adjacent layer, and therefore will generate a net spin current $\bm{J}_s \parallel \Delta T \ \hat{\mathbf{n}}$ with polarization $\bm{\sigma}$ parallel to the spin direction at the uncompensated monolayer \cite{BenderPRL}, which is $\bm{s}_{top}$ if the Pt is above the NiO and $\bm{s}_{bottom}$ if the Pt is beneath the NiO. At the compensated interface, the symmetry between magnon modes is preserved, and an interfacial thermal gradient should produce no net spin current. Although the uncompensated interface in the schematic in Fig.~2(a) is atomically flat, the presence of roughness in real samples does not alter the interpretation of the AF LSSE signal as long as the lateral length scale of height variations is much smaller than the laser spot diameter (see the supporting information \footnotemark[1] for more details).

Our experimental test of interfacial AF LSSE is shown in Fig.~2(b) and (c). We take AF LSSE images of MBE-grown $\mathrm{MgAl}_2\mathrm{O}_4$/165 nm NiO(111)/Pt and MgO(001)/136 nm NiO(001)/Pt, which have uncompensated and compensated interfaces, respectively. Both samples are patterned into 20 $\mu$m-wide Hall bars with similar growth conditions and sample resistances. We find that AF LSSE images of NiO(001) show negligible signal compared with NiO(111), which is consistent with the picture of the interfacial AF LSSE and also suggests that bulk AF LSSE \cite{HolandaAPL, RezendePRB2} does not significantly contribute to our signal. This interpretation is further supported by finite-element simulations, discussed in the supporting information, which show that the laser-induced thermal profile is dominated by temperature drops at the Pt/NiO interfaces rather than a temperature gradient in the NiO bulk.

\begin{figure*}[!htb]
\centering
\includegraphics[scale=0.39]{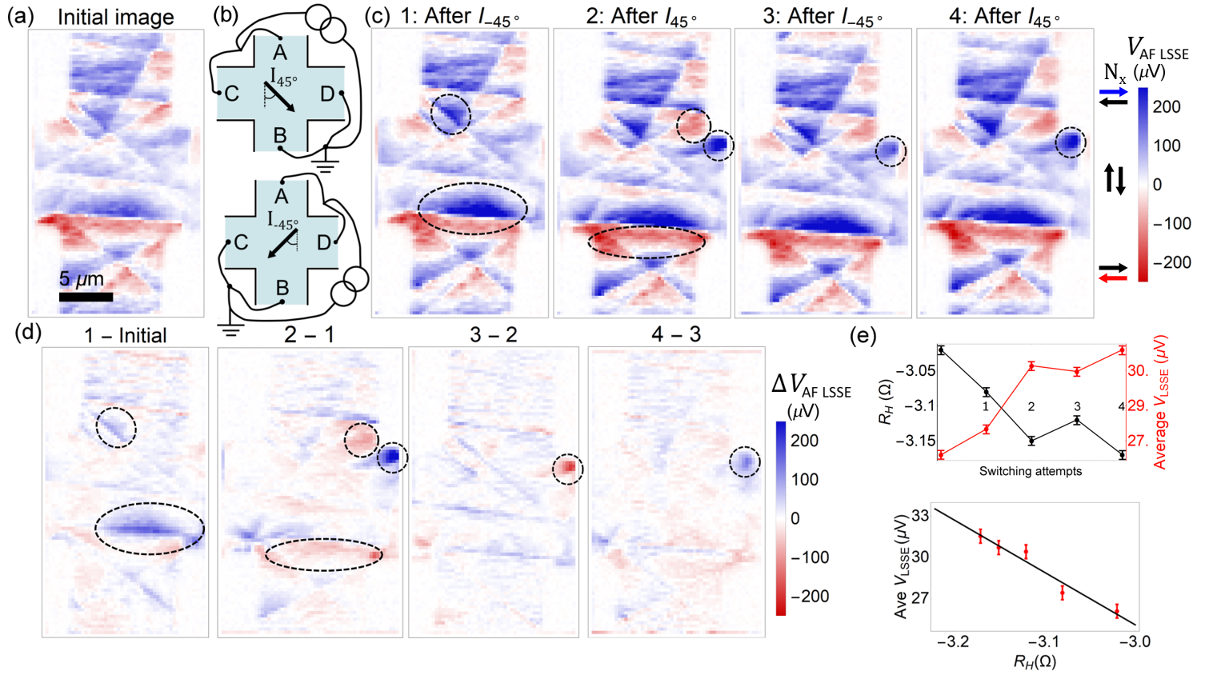}
\caption{AF LSSE imaging of spin-torque switching in a 4 nm Pt/6 nm NiO(111)/4 nm Pt trilayer. (a) The initial image, taken before applying current. (b) Schematic of the writing process. We apply current to two arms of the Hall cross such that the current density in the center flows along 45$^\circ$ diagonals. (c) Imaging while toggling between $I_{45^\circ}$ and $I_{-45^\circ}$. Highlighted in dashed line are changes in contrast at the top right corner, where the current density is greatest, and near T-domain boundaries, where the spins are less strongly exchange-coupled and more likely to switch. (d) Successive differences between the AF LSSE images in (a), showing the domains that switch more clearly. Both positive and negative contrast in difference images may reflect domains rotating in opposite directions. (e) $R_H$ and the integrated AF LSSE signal, $\langle V_{AF~LSSE} \rangle$, measured while toggling between $I_{45^\circ}$ and $I_{-45^\circ}$. Values corresponding to the images shown are labeled. (f) $\langle V_{AF~LSSE} \rangle$ plotted as a function of SMR, measured through changes in the Hall resistance $R_H$. The near-linear correlation shows the small-angle correspondence between $\langle V_{AF~LSSE} \rangle$ and $R_H$.}  
\end{figure*}

We can further distinguish between an AF LSSE at an uncompensated interface and a possible FM LSSE originating from pinned uncompensated moments (UMs). These UMs could arise from interfacial roughness \cite{CharilaouJAP} or defects, both at the interface and in the bulk of the AF \cite{SchullerJMMM}, and would be detectable by other magnetometry techniques. Therefore, we perform scanning SQUID microscopy at 4 K to search for microscopic moments, and polarized neutron reflectometry at room temperature to detect a global moment at 1 T applied field. Both sets of measurements were made on sputtered samples and are described in more detail in the supporting information \footnotemark[1]. We find no magnetic moment within sensitivity, which places an upper limit of $8 \times 10^{-4} \ \mu_B$/Ni on the local moment that could be present. From this value we calculate a maximum bulk magnetization of 110 $\mathrm{A}/\mathrm{m}$ in our NiO, three orders of magnitude less than the bulk magnetization in YIG at room temperature. Based on these measurements we rule out a FM LSSE and conclude that the AF LSSE signal in NiO arises from the uncompensated interface.

\section{Imaging spin-torque switching in P\lowercase{t}/N\lowercase{i}O/P\lowercase{t} trilayers}
 
Having established a mechanism for the signal contrast, we move onto image current-induced spin-torque switching in Hall crosses, initially following the procedure in Refs. \cite{MoriyamaArxiv} and \cite{ChenPRL}. We apply DC writing current and characterize the N\'eel state electrically, using the antiferromagnetic analog of spin Hall magnetoresistance (SMR) \cite{NakayamaPRL, HoogeboomAPL, FischerPRB} by measuring the change in the Hall resistance $R_H$: $R_H = -\Delta R_{SMR} \sin \theta \cos \theta$, where $\theta$ is the angle between the spatially-averaged N\'eel vector $\bm{N}$ and the reading current $\bm{J}_R$. To maximize $\Delta R_H$, we apply writing current to two of the arms of the Hall cross such that the current density in the center of the device flows along $\pm$45$^\circ$ (schematically illustrated in Fig.~3(b)). Using a finite-element simulation described further in the supporting information \footnotemark[1], we estimate a writing current density of $8.0 \times 10^7 \ \mathrm{A}/\mathrm{cm}^2$ at the corners and $3.1 \times 10^7 \ \mathrm{A}/\mathrm{cm}^2$ in the center of the cross. Hereafter we refer only to the density in the center of the cross, $J_W$. After each application of writing current, we measure the Hall resistance $R_H$ by applying a reading current density $J_R = 1.5\times 10^6 \ \mathrm{A}/\mathrm{cm}^2$ from A to B and measuring the voltage from C to D in Fig.~3(b).

We initially employ epitaxial sputtered 4 nm Pt/6 nm NiO(111)/4 nm Pt trilayers, following the argument of Ref. \cite{MoriyamaArxiv} (further demonstrated in Ref. \cite{MoriyamaPRL} in synthetic antiferromagnets) that spin torque at both the top and bottom interface of the NiO leads to more coherent rotation of the N\'eel orientation throughout the AF thickness. There is a potential complication interpreting the AF LSSE images in trilayers because both Pt/NiO interfaces can contribute to the signal. Separately comparing AF LSSE images of single-interface Pt/NiO and NiO/Pt samples (shown in the supporting information) indicates that in the sputtered samples, the signal from the bottom interface dominates, which may reflect higher interface quality at the bottom than the top Pt/NiO interface. This Pt/NiO/Pt trilayer has also been annealed at 200 $^\circ$C for 20 min to increase the S-domain size, however, we also find that annealing decreases domain wall motion as described in the supporting information \footnotemark[1].

AF LSSE images of a 4 nm Pt/6 nm NiO/4 nm Pt trilayer before and after four sequential applications of $J_W = 3.1 \times 10^7 \mathrm{A}/\mathrm{cm}^2$ are shown in Fig.~3(b), with the writing current direction alternating between $+45^\circ$ and $-45^\circ$. Although most domains are unaltered, we observe changes in contrast (highlighted inside the black dashed enclosures) at a sample corner, where the current density is highest, and at T-domain boundaries, where the spins are less strongly exchange-coupled.  To quantify these changes to the N\'eel orientation, we calculate sequential differences between images, shown in Fig.~3(d). We observe both positive and negative changes in contrast in different parts of the sample, which could be due to different S-domains rotating in opposite directions, as seen in imaging studies of switching in CuMnAs \cite{GrzybowskiPRL}. At the current density used, we estimate that the maximum Oersted field is $\sim10$ mT. We show in the supporting information \footnotemark[1] that the AF LSSE signal is unaffected by field up to $\pm$250 mT, which rules out the Oersted field from the writing current as the mechanism responsible for switching.

 To compare AF LSSE imaging with electrical measurements of N\'eel order using $R_H$, we take the average of all the pixels in and near the cross center in each image (described in the supporting information \footnotemark[1]) to obtain the integrated AF LSSE signal, $\langle V_{AF~LSSE}\rangle$. Although $\langle V_{AF~LSSE} \rangle$ and $R_H$ are both measures of the average N\'eel orientation in the cross center, they have different symmetries: $R_H \propto \cos \theta \sin \theta$, where $\theta$ is the angle between the average N\'eel vector and the SMR reading current $\bm{J}_R$, while $\langle V_{AF~LSSE} \rangle \propto \cos \phi$, where $\phi$ is the angle between the average N\'eel vector and the voltage contacts. Since we apply $\bm{J}_R$ along $x$ in this device, here $\theta = \phi-\pi/2$ and $\langle V_{AF~LSSE} \rangle \propto -\sin \theta$. In this sample, most of the changes in contrast occur where $\bm{N}$ appears to be nearly saturated in the $+x$-direction, so that locally $\theta \approx 0$. Near $\theta = 0$, $\sin \theta \cos \theta \approx \sin \theta$. Therefore, $\langle V_{AF~LSSE} \rangle$ tracks $R_H$ point-by-point, shown in Fig.~3(e). Plotting one versus the other yields a near-linear correlation with negative slope, shown in Fig. 3(f) with a linear fit drawn as a guide to the eye. This correspondence indicates that changes in contrast indeed represent antiferromagnetic switching.

We expect that AF switching can occur either by domain rotation, which would manifest in the AF LSSE images as changes of contrast level within S-domains, or by domain wall motion. The switching in Fig.~3 manifests as changes in contrast within domains while domain walls remain stationary within the resolution limit, which indicates domain rotation. In this sample we observe changes in color shade but not changes in sign of $N_x$ (blue to red or vice versa), which indicates that $N$ rotates by acute angles. Although we cannot obtain an absolute angle of rotation, we can obtain a lower bound by taking the maximum and minimum $V_{AF~LSSE}$  to correspond to $\theta = 90^\circ$ and $-90^\circ$, respectively. In this case we estimate that the average N\'eel vector at the corner rotates 22$^\circ$ between images \textit{1} and \textit{2} in Fig.~3(c), and 10$^\circ$ between images \textit{2} and \textit{3}. 

Previous studies of magnetic field-induced domain rotation in 120 nm-thick NiO \cite{FischerPRB} modeled switching as 120$^\circ$ flopping between $\langle 11 \bar{2} \rangle$ in-plane easy axes. In our samples, however, the S-domains have random in-plane orientation, which is consistent with XMLD-PEEM images of similar Pt/NiO/Pt trilayers \cite{MoriyamaArxiv}. This is consistent with an increased role of magnetoelastic stress in our 6 nm-thick samples, which favors a multidomain state with zero average strain. While the effective field from in-plane anisotropy is $H_{Az} =$ 11 mT in bulk NiO \cite{RezendePRB}, the destressing field reported in 120 nm-thick NiO in Ref.~\cite{FischerPRB} is 46 mT. We expect the destressing field to be even higher in 6 nm NiO. Therefore, because the spins are not restricted to the $\langle 11 \bar{2} \rangle$ axes in our samples, they can switch by continuous in-plane rotation. 

\section{Resolving domain rotation and domain wall motion in M\lowercase{g}O/P\lowercase{t}/N\lowercase{i}O bilayers}
 
After correlating the AF LSSE images with electrical readout of the N\'eel order through SMR in Pt/NiO/Pt trilayers, we move on to imaging switching in the sputtered NiO(111)/Pt bilayer from Fig.~1 after applying current along the device channel. The bilayer does not have the potential difficulty of superposing signal from two Pt/NiO interfaces. Furthermore, applying current along the device channel yields more uniform current density, leading to larger-scale, more easily resolvable changes in image contrast. Fig.~4(a) shows AF LSSE images before switching, and then after applying progressively greater current densities, from $5.0 \times 10^7 \ \mathrm{A}/\mathrm{cm}^2$ at 20 mA to  $1.1 \times 10^8 \ \mathrm{A}/\mathrm{cm}^2$ at 42 mA, first at positive polarity (flowing down) and then negative polarity (flowing up). Prominent regions of switching are highlighted in the black enclosure as a guide to the eye.

\begin{figure*}
\centering
\includegraphics[scale=0.78]{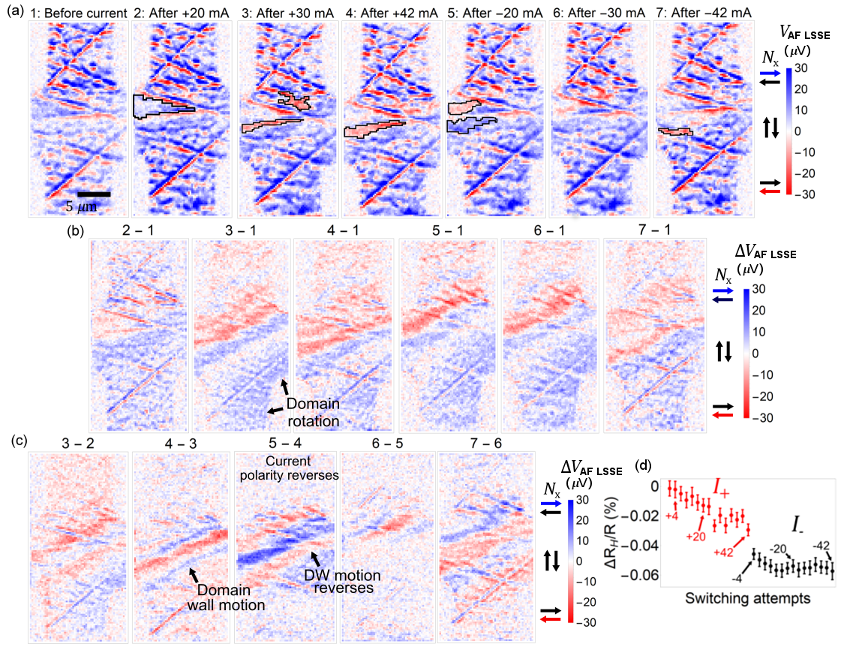}
\caption{Switching in the MgO/Pt/NiO bilayer from Fig. 1 after applying current along the channel direction. (a) AF LSSE images, before applying current and then after applying first positive (down) and negative (up) current. Some prominent regions of switching are highlighted in black line, including domain growth after +42 mA and domain wall motion after -20 mA. (b) Cumulative differences from the initial state (1). Large-scale patterns of nearly uniform positive contrast in the lower half of the device show different S-domains rotating by the same in-plane angle. (c) Sequential differences between the AF LSSE images in (a). We observe domain wall motion after +42 mA that reverses direction when the current polarity is reversed, consistent with theoretical predictions of a chiral domain wall force $\bm{F}_{DW}$. (d) SMR measurements of the average N\'eel orientation for a similar MgO/Pt/NiO bilayer after applying the same currents, first positive and then negative (labeled in mA). $R_H$ does not depend on current polarity, which is consistent with domain wall motion due to $\bm{F}_{DW}$.}
\end{figure*}

Because the switching in Fig.~4 is spatially distributed and nonuniform, and AF LSSE and SMR have different symmetries, we cannot correlate the AF LSSE signal with SMR like we do in Fig.~3. Therefore, we compare our AF LSSE images to theoretical models of switching. Refs. \cite{MoriyamaArxiv} and \cite{ChenPRL} model switching in the high-current limit as coherent rotation of spins within an S-domain (domain rotation), which we observe in Fig.~3. The model in Ref. \cite{BaldratiArxiv} distinguishes three separate switching mechanisms, with predictions summarized as follows:

(1) The out-of-plane component of the spin torque rotates spins within the easy plane, rotating all S-domains by the same angle.

(2) The in-plane component of the spin torque creates an additional effective anisotropy, resulting in a translational ponderomotive force $\bm{F}_{pond}$ proportional to $J_W^2$ on the S-domain wall. $\bm{F}_{pond}$ rotates $\bm{N} \parallel \bm{J}_W$ and is current polarity-independent.  

(3) The spin torque directly rotates the spins within the domain walls, leading to a chiral domain wall force $\bm{F}_{DW}$ that goes as $\bm{J}_W$. $\bm{F}_{DW}$ can rotate $\bm{N}$ either towards or perpendicular to $\bm{J}_W$, depending on the domain wall chirality, therefore it should have no net effect on $R_H$ with randomly oriented S-domains. $\bm{F}_{DW}$ should also reverse direction when $\bm{J}_W$ reverses.

To quantitatively characterize the switching, we take AF LSSE image differences as in Fig.~3. We take cumulative differences from the initial state in Fig.~4(b), which show polarity-independent switching, and sequential differences between adjacent pairs of AF LSSE images in Fig.~4(c), which better illustrate polarity-dependent switching. After applying 30 mA (seen in image \textit{3~-~1}) and in subsequent difference images in Fig.~4(b), we see large-scale, nearly uniform positive (blue) contrast in the lower half of the cross, labeled \textit{domain rotation}. More uniform contrast in the difference images than in the AF LSSE images themselves indicate that different S-domains are rotating by the same angle, consistent with the out-of-plane spin torque in mechanism 1 in Ref.~\cite{BaldratiArxiv}. This domain rotation saturates after +30 mA ($7.5\times10^7~\mathrm{A}/\mathrm{cm}^2$), and does not reverse when the current polarity reverses. We primarily observe $|N_x|$ increasing - blue domains become more blue and red domains more red - which means $\bm{N}$ rotates $\perp \bm{J}_W$. 

From images \textit{3} and \textit{4} in the AF LSSE images in Fig.~4(a), we resolve switching by domain wall motion, which appears as a negative (red) horizontal stripe in the sequential \textit{4 - 3} image in Fig.~4(c). Domain rotation and domain wall motion occurs in response to writing current as low as 4 mA ($1.0\times10^7~\mathrm{A}/\mathrm{cm}^2$). Interestingly, we find that the domains continue to move for 2-3 hours after the current is turned off (shown in the supporting information \footnotemark[1]), which may be due to magnetoelastic stresses causing subthreshold domain wall creep after the spin torque rotates the S-domains out-of-equilibrium. Although the domain configuration after +42 mA in \textit{4} does not creep in time, domain wall motion reverses after applying -20 mA, seen in \textit{5 - 4}, and subsequently almost ceases, seen by weaker contrast in \textit{6 - 5} after applying -30 mA. DW motion that reverses when the current polarity is reversed points to the chiral force $\bm{F}_{DW}$ as the origin. In Fig.~4(d) we show SMR measurements of a similar MgO/Pt/NiO cross while applying the same currents, first positive and then negative. We find $R_H$ does not depend on current polarity, which is consistent with the prediction that the effects of $\bm{F}_{DW}$ would not be reflected by changes in $R_H$. 

Summarizing our results, we identify both domain rotation and domain wall motion acting simultaneously, which are consistent with the out-of-plane spin torque and chiral domain force, respectively, described in Ref.\cite{BaldratiArxiv}. At the current densities applied, from $1.0 \times  10^7~\mathrm{A}/\mathrm{cm}^2$ to $1.0 \times 10^8~ \mathrm{A}/\mathrm{cm}^2$, we do not observe $\bm{N}$ rotating towards $\bm{J}_W$ from the ponderomotive force $\bm{F}_{pond}$, which is expected to dominate at higher current densities (above $7-9 \times 10^7~\mathrm{A}/\mathrm{cm}^2$, depending on the strain).  Further imaging studies on thicker NiO samples may be required to observe $\mathbf{F}_{pond}$. 
 
 Our results complement the XMLD-PEEM images of switching in Refs. \cite{MoriyamaArxiv} and \cite{BaldratiArxiv}. Although Ref. \cite{MoriyamaArxiv} shows domain wall motion and Ref. \cite{BaldratiArxiv} appears to show domain rotation in response to current, distinguishing several simultaneously acting switching mechanisms requires systematic repeated imaging of multiple samples, which may not be practical with limited beam time at a synchrotron facility. 

\section{Conclusion} 
 
In conclusion, we demonstrate interfacial AF LSSE as the basis for a powerful tabletop technique for imaging in-plane N\'eel order in an AF insulator. This magneto-thermal microscope uses equipment that is readily available in many laboratories, thus enabling in-depth and high-throughput studies of AF spintronics, which was previously limited by the availability of a few coherent x-ray facilities.  Using this capability, we probe the microscopic behavior of spin torque switching of N\'eel order in Pt/NiO/Pt trilayers and Pt/NiO bilayers. We find that switching occurs by domain rotation and domain wall motion acting simultaneously, and that magnetoelastic stresses play an important role in determining both the equilibrium domain structure and the fraction of domains that switch. These insights provide critical understanding of spin torque switching in NiO and point the way towards systematic optimization of antiferromagnetic spintronic devices. Moreover, we expect AF LSSE microscopy to extend to a wide variety of antiferromagnetic insulators with uncompensated interfaces, which can aid in the development of new device technologies based on different antiferromagnets.

\begin{acknowledgments}
We thank Rembert Duine, Yaroslav Tserkovnyak, Jason Bartell, Emrah Turgut, and Farhan Rana for useful discussions. This research was supported by the Cornell Center for Materials Research with funding from the NSF MRSEC program (DMR-1719875) and by JSPS KAKENHI Grant Numbers JP15H05702, JP17H04924, and JP17H05181. This work made use of the CCMR Shared Facilities and the Cornell NanoScale Facility, an NNCI member supported by NSF Grant ECCS-1542081. N.S. acknowledges National Science Foundation [Platform for the Accelerated Realization, Analysis, and Discovery of Interface Materials (PARADIM)] under Cooperative Agreement No. DMR-1539918 and Cornell University Center for Advanced Computing for his time at Cornell University.
\end{acknowledgments}

\clearpage
\widetext
\begin{center}
\textbf{\large Supplemental Information for ``Spin Seebeck imaging of spin-torque switching in antiferromagnetic Pt/NiO heterostructures"}
\end{center}

\setcounter{equation}{0}
\setcounter{figure}{0}
\setcounter{table}{0}
\setcounter{page}{1}
\setcounter{section}{0}
\makeatletter

\renewcommand{\thepage}{S\arabic{page}}
\renewcommand{\thesection}{S\arabic{section}}
\renewcommand{\thetable}{S\arabic{table}}
\renewcommand{\thefigure}{S\arabic{figure}}

\section{Epitaxial growth of P\lowercase{t}/N\lowercase{i}O and P\lowercase{t}/N\lowercase{i}O/P\lowercase{t}} 

\begin{figure}[htb]
\centering
\includegraphics[scale=0.3]{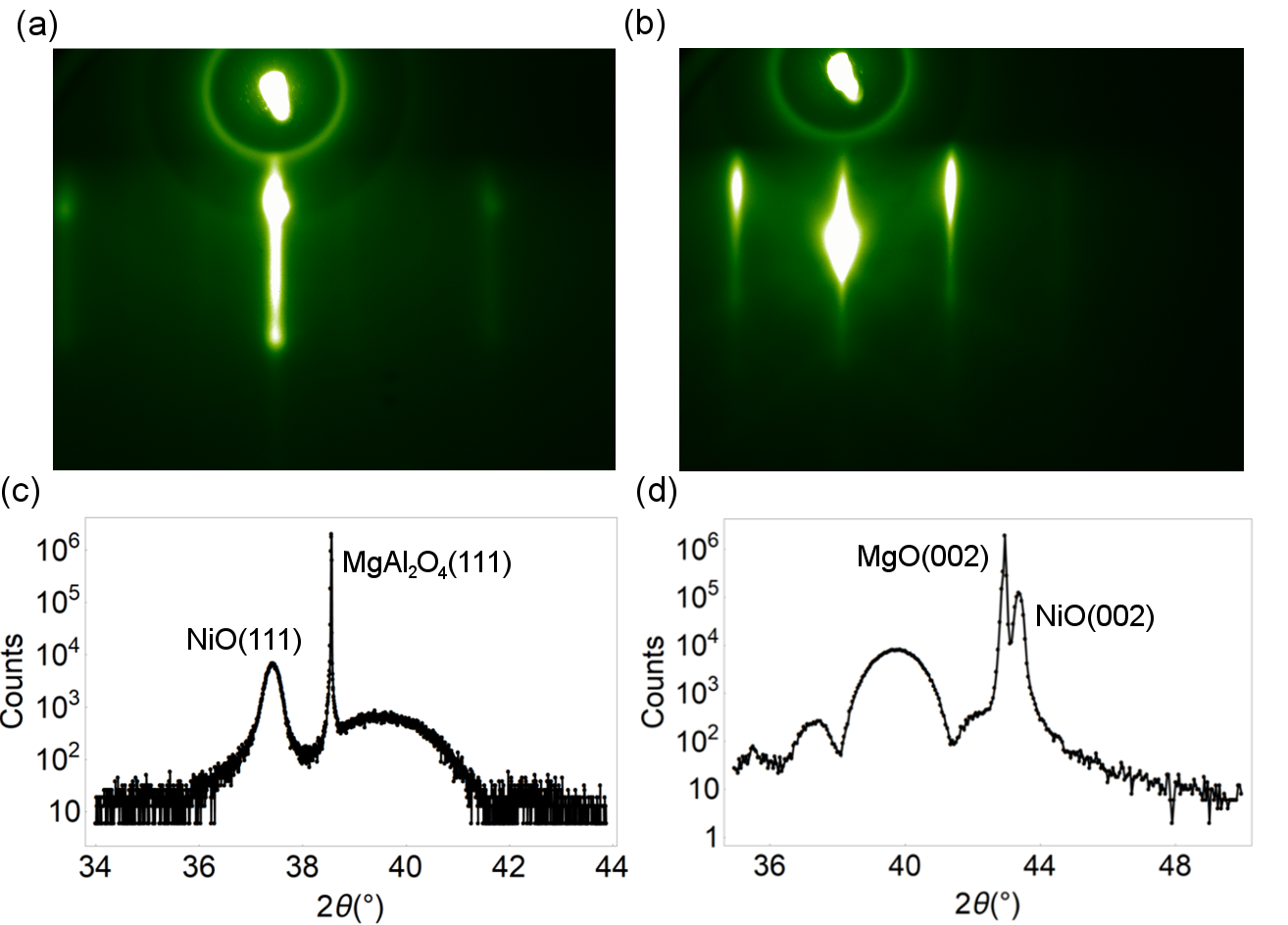}
\caption{(a,b) RHEED images of the uncompensated NiO(111) and compensated NiO(001) surfaces from the MBE-grown samples in Fig.~2 of the main text. (c,d) The corresponding $\theta-2\theta$ scans, showing epitaxial (111) and (001) growth. }
\label{MBE growth}
\end{figure}

The uncompensated and compensated Pt/NiO bilayers in Fig.~2 of the main text are $\mathrm{MgAl}_2\mathrm{O}_4$(111)/136 nm NiO(111)/6 nm Pt and MgO(001)/168 nm NiO(001)/6 nm Pt, respectively, grown by molecular-beam epitaxy (MBE). The $\mathrm{MgAl}_2\mathrm{O}_4$(111) is annealed at 900 $^\circ$C before depositing NiO(111) at 50 $^\circ$C, and the MgO(001) is annealed at 700 $^\circ$C before depositing NiO(001 at 500 $^\circ$C. RHEED images of both the NiO(111) and NiO(001) surface as well as the $\theta-2\theta$ scans are shown in Fig.~\ref{MBE growth}. 

\begin{figure}[htb]
\centering
\includegraphics[scale=0.23]{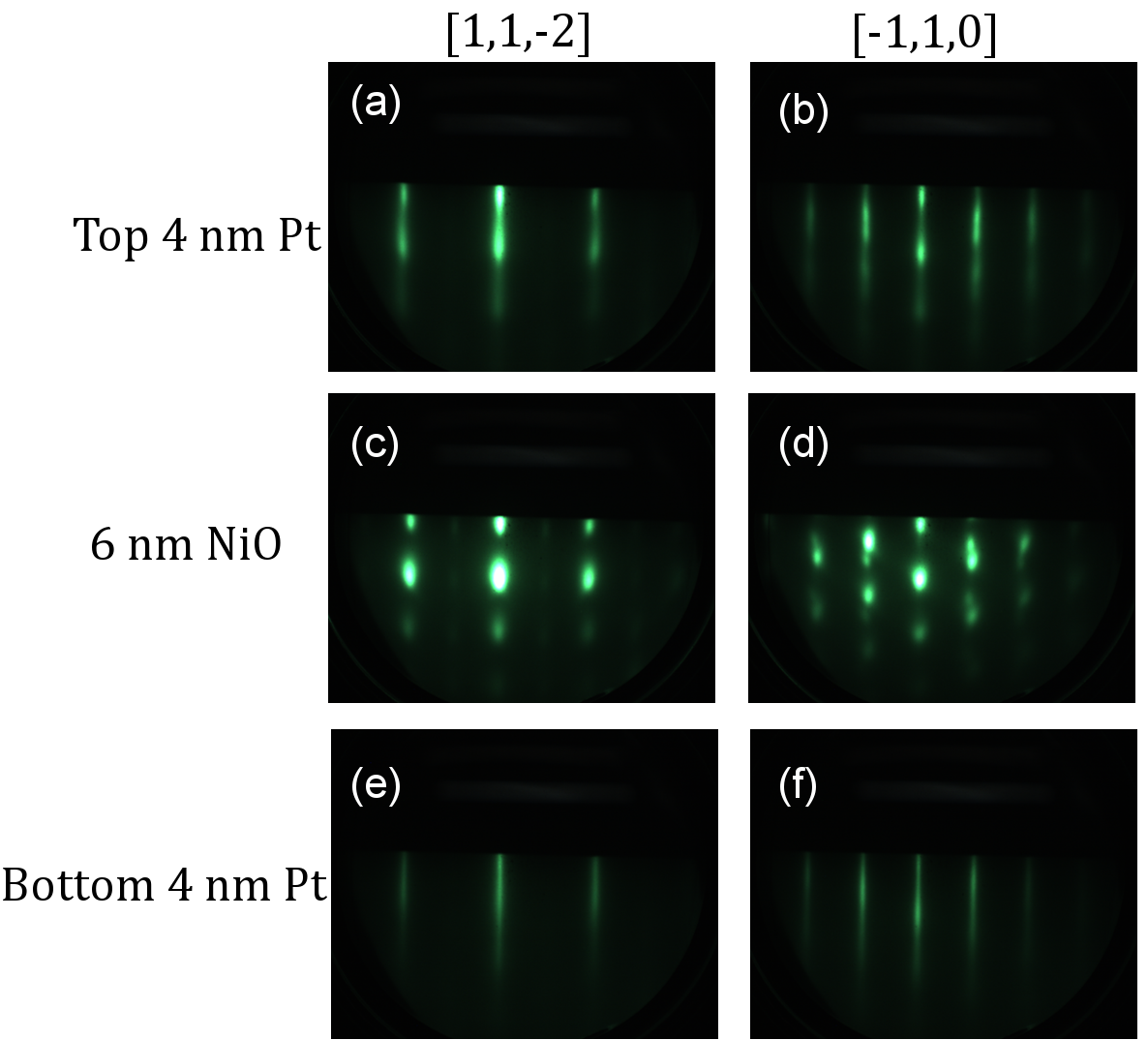}
\caption{RHEED images showing epitaxial (111) growth at the top interface of the top Pt layer (a,b), the top Pt/NiO interface (c,d), and the bottom Pt/NiO interface (e,f). }
\label{Sputtered RHEED}
\end{figure}

All other NiO/Pt and Pt/NiO/Pt films in the main text and supporting information are grown by magnetron sputtering at room temperature on MgO(111) single-crystal substrates, similar to the samples in Ref. \cite{MoriyamaArxiv}. Figure~\ref{Sputtered RHEED} shows the RHEED images for each interface in the Pt/NiO/Pt trilayer, which indicate epitaxial (111) crystal orientation for each layer.

\section{Experimental details of AF LSSE microscopy}

\begin{figure}[htb]
\centering
\includegraphics[scale=0.13]{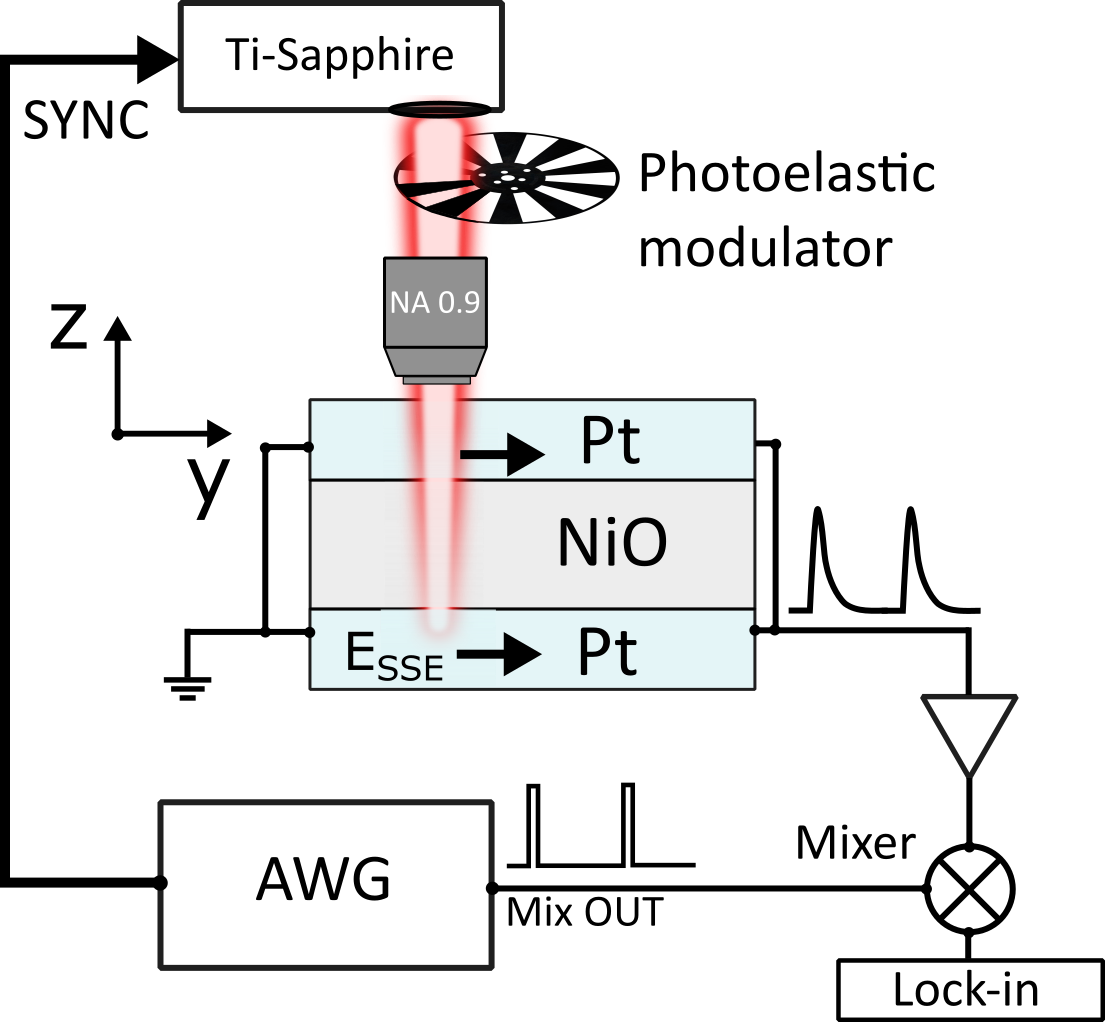}
\caption{Schematic of the measurement circuit. 3 ps pulses from a Ti:Sapphire laser at 76 MHz repetition rate generate voltage pulses, which are amplified and mixed with a reference voltage pulse train from an arbitrary waveform generator (AWG). We modulate the laser intensity at 100 kHz with a photoelastic modulator (PEM) and detect the resulting voltage with a lock-in amplifier. }
\label{schematic}
\end{figure}

Our measurement circuit, schematically shown in Figure~\ref{schematic}, is similar to that described in \cite{TRANENatComm} and \cite{TRANEPRAppl2}. We generate local heating with 3 ps-wide pulses from a Ti:Sapphire laser, at a repetition rate of 76 MHz, which we focus down to a ~650 nm diameter spot. The laser pulse train produces a spin Seebeck voltage pulse train, $V_{AF~LSSE}$, at the same repetition rate. The duration of each $V_{AF~LSSE}$ voltage pulse is equal to the time duration of the interfacial temperature drop $\Delta T$ \cite{TRANENatComm}, which we estimate from the finite-element simulations in Fig.~\ref{bilayer simulations} and Fig.~\ref{trilayer simulations} to be ~50 ps. We feed the pulses into a 50 $\Omega$ coplanar waveguide transmission line. After amplifying the pulses we perform homodyne detection by electrically mixing the pulses with a train of 600 ps-wide square pulses produced by an arbitrary waveform generator (AWG). Longer mixing pulses yield larger signal, but if the mixing pulses are too long the signal-to-noise decreases because we are more exposed to noise between thermal pulses. Through experience we have found 600 ps width to be a good compromise. We synchronize the laser repetition rate to the external AWG pulse train frequency using a Coherent Synchro-lock AP 9th-harmonic locking mechanism inside the laser cavity. To take advantage of low-noise lock-in detection techniques, we modulate the intensity of the laser beam at 100 kHz with a photoelastic modulator and a polarizer.

Note that impedance mismatch between the sample resistance and the 50 $\Omega$ transmission line results in an overall signal scaling factor that depends on sample resistance, which ranges between 450 $\Omega$ and 650 $\Omega$ in our Hall crosses. To remove this dependence, all AF LSSE images, both in the main text and in the supplementary information, have been normalized by the factor $(R_{sample} + 50 \ \Omega) /50 \  \Omega$. 

\section{Characterization of the AF LSSE signal}

\subsection{Note on annealing of NiO samples}

S-domains in NiO become larger and more uniform with heat treatment \cite{AraiPRB}. We present two kinds of samples in this work: unannealed samples that have undergone no heat treatment, and samples that have been annealed at 200 $^\circ$C for 20 min. In Fig.~3 of the main text, we show switching of an annealed sample because the spatially localized rotation is easier to interpret and correlates better with SMR measurements. The other samples in the main text are unannealed. While unannealed samples have 2-5 $\mu$m S-domain size and are better resolved compared to the submicron domain size in unannealed samples, we find unannealed samples have a factor of 5 greater switching efficiency measured through SMR as well as greater and more spatially widespread changes in AF LSSE images. This can be seen by comparing AF LSSE and SMR measurements of switching in unannealed Pt/NiO/Pt in Fig.\ref{multiple switching} and annealed Pt/NiO/Pt in Fig. 3 of the main text, both performed at $I_{\pm 45^{\circ}}$ with the same $4.1 \times 10^7~\mathrm{A}/\mathrm{cm}^2$ current density.

\subsection{Effects of heating and laser fluence on S-domain structure}

\begin{figure}[htb]
\centering
\includegraphics[scale=0.36]{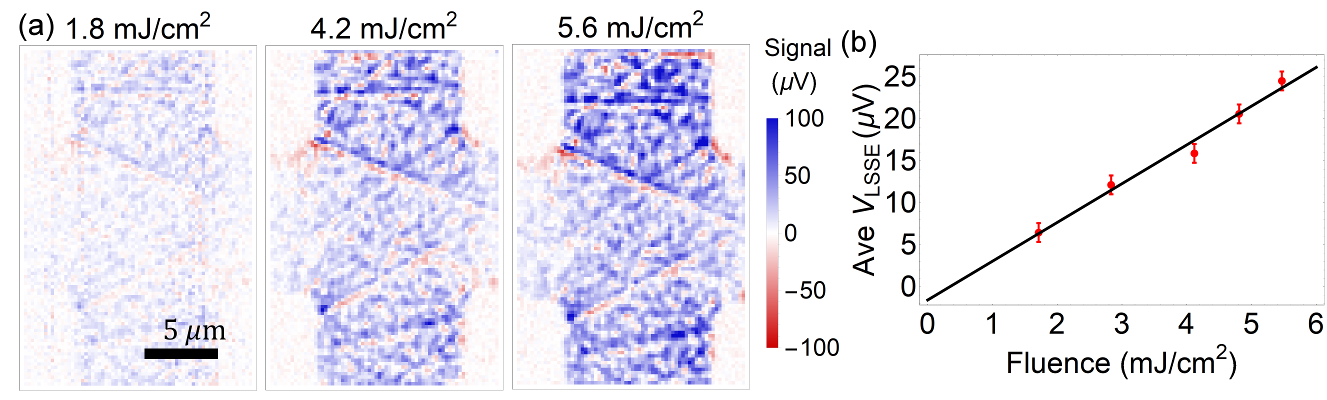}
\caption{(a) AF LSSE images of an unannealed 10 $\mu$m-wide Pt/NiO/Pt Hall cross at 3 different laser fluences. Below $5.6 \  \mathrm{mJ}/\mathrm{cm}^2$, the antiferromagnetic domain structure is not affected by laser heating. (d) The average AF LSSE image signal as a function of fluence, showing that the signal scales linearly with fluence.}
\label{fluence dependence}
\end{figure}

We perform AF LSSE imaging on an unannealed 10 $\mu$m-wide Pt/NiO/Pt trilayer sample as a function of laser fluence, first to check linearity of the signal at low fluence and second to determine the effect of laser-induced heating on the N\'eel order at high fluence. Fig.~\ref{fluence dependence}(a) shows that from 1.8 $\mathrm{mJ}/\mathrm{cm}^2$ to $5.6 \ \mathrm{mJ}/\mathrm{cm}^2$, the S-domain structure is unaffected, and Fig.~\ref{fluence dependence}(b) shows that the integrated AF LSSE signal over the whole sample scales linearly with fluence. Since the images in the main text are taken at $3.4 \ \mathrm{mJ}/\mathrm{cm}^2$ fluence, this result suggests that in those images, the laser probes the N\'eel order without perturbing it.

\begin{figure}[htb]
\centering
\includegraphics[scale=0.38]{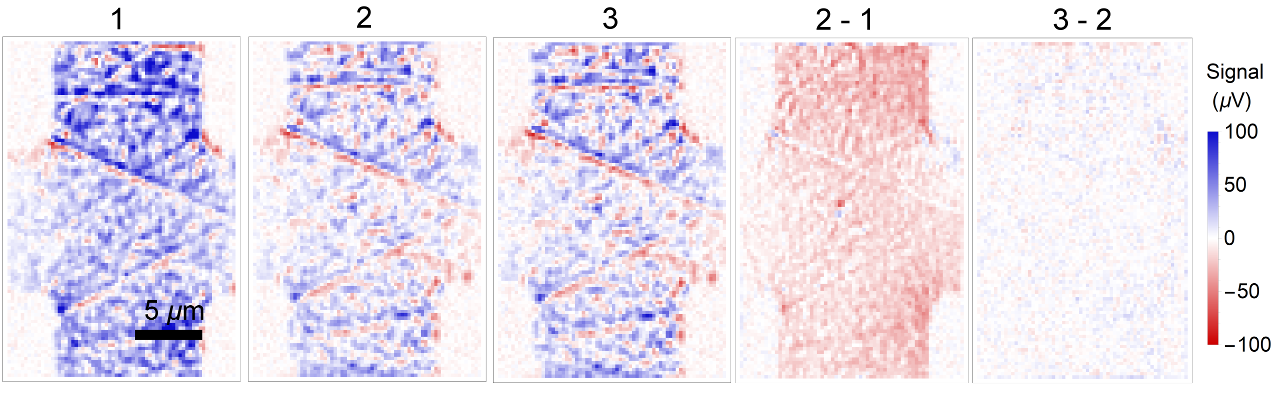}
\caption{Repeated imaging of an unannealed 10 $\mu$m-wide Hall cross at 5.6 $\mathrm{mJ}/\mathrm{cm}^2$ fluence, which results in an estimated maximum $\Delta T=90$ K at the top NiO surface. The image is initially dominated by positive $N_x$ projection in $\mathbf{1}$. Laser heating locally destabilizes the S-domain orientations, causing local flopping of the N\'eel orientation as seen in $\mathbf{2}$ and $\mathbf{2 - 1}$. Once the domains reach their local lowest-energy orientations, they are unaffected by further laser heating, as shown in $\mathbf{3}$ and $\mathbf{3 - 2}$. }
\label{change from fluence}
\end{figure}

We then image repeatedly at 5.6 $\mathrm{mJ}/\mathrm{cm}^2$ laser fluence, which heats the top NiO surface to a maximum of 390 K as estimated from finite-element calculations. This is comparable to the temperature change that we estimate from applying a writing current density of $4\times10^7 \mathrm{A}/\mathrm{cm}^2$, so it is reasonable to think the laser may affect the local N\'eel orientation. In Fig.\ref{change from fluence} we plot the initial image ($\mathbf{1}$), an image after scanning the laser over the sample 4 times without applying current ($\mathbf{2}$), another image taken immediately afterwards ($\mathbf{3}$), and the image differences ($\mathbf{2 - 1}$ and $\mathbf{3 - 2}$).

\begin{figure}[htb]
\centering
\includegraphics[scale=0.38]{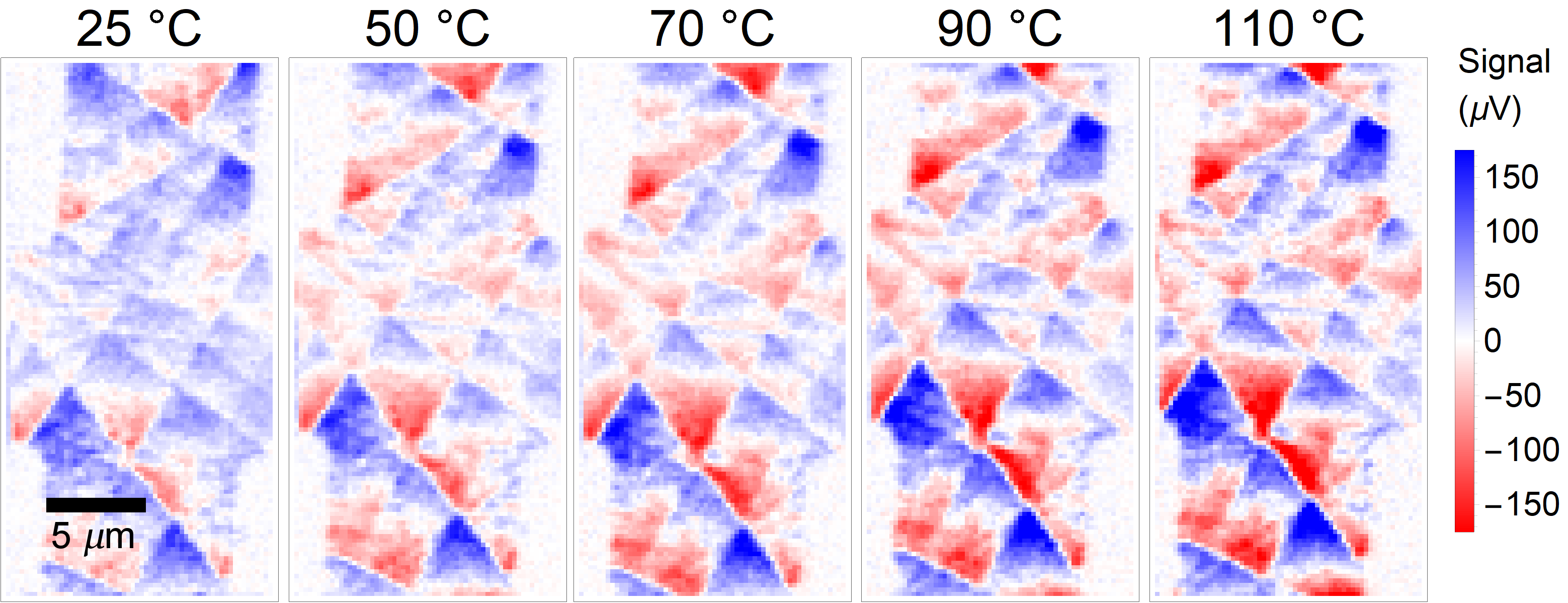}
\caption{AF LSSE imaging of a Pt/NiO/Pt trilayer as a function of background temperature. Heat treatment increases S-domain size, randomizes the S-domain orientation, and makes the spins more uniform within the S-domains.}
\label{temp dependence}
\end{figure} 

We observe that the S-domains initially have mostly positive $N_x$ in $\mathbf{1}$. After repeatedly scanning the laser, we see in $\mathbf{2}$ that the dominant positive orientation has disappeared but the T-domain structure is unchanged. $\mathbf{3}$ is nearly pixel-for-pixel identical to $\mathbf{2}$, which means that the laser has no further effect after $\mathbf{2}$. Laser heating locally destabilizes the N\'eel order until the S-domains are stable in local lowest-energy configurations, which is why the orientation is more random in $\mathbf{2}$ and repeats almost exactly in $\mathbf{3}$. This process is irreversible and occurs almost uniformly over the whole sample (seen in $\mathbf{2 - 1}$), quite different from the localized domain rotation and domain wall motion produced by spin-torque switching.

We further investigate the effect of background heating on the N\'eel orientation by imaging an annealed trilayer with larger S-domains as a function of temperature in Fig.~\ref{temp dependence}. The T-domain orientation is unchanged upon heating, as we expect. Although the overall S-domain orientation is randomized, the spins within the S-domain become more uniform upon heating, in agreement with previous XMLD-PEEM imaging studies of annealing \cite{AraiPRB}. 

\subsection{Characterizing T-domains with atomic force microscopy}

\begin{figure}[!htb]
\centering
\includegraphics[scale=0.5]{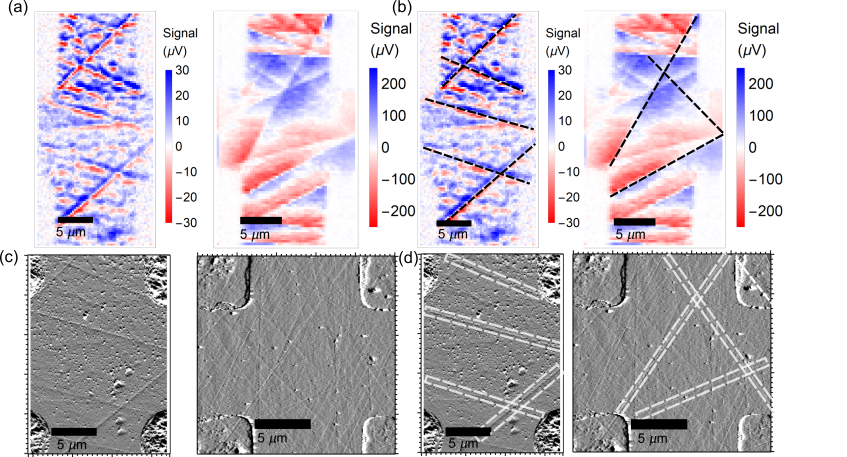}
\caption{Comparing T-domain boundaries in AF LSSE images, visible as sharp diagonal lines, to atomic force microscopy (AFM) images. (a) AF LSSE images of the MgO/Pt/NiO bilayer in Fig.~1 and 4 of the main text and a Pt/NiO/Pt trilayer. (b) Corresponding AFM amplitude images. T-domain lines in the AFM also appear in the AF LSSE image in the bilayer, while many T-domain lines in the AFM image of the trilayer do not appear in the AF LSSE image. (c,d) The same AF LSSE and AFM images with dashed lines to highlight T-domain boundaries that are seen with both techniques.}
\label{AFM comparison}
\end{figure}

In addition to antiferromagnetic S-domain contrast, we observe sharp straight lines of contrast in the AF LSSE images that represent nonmagnetic T-domain walls between crystal grains. Within a T-domain, the sample surface is thermally isotropic in the plane, therefore conventional Seebeck voltages from laser-induced in-plane thermal gradients cancel in detail. At a T-domain boundary, however, there is a discontinuity in in-plane thermal conductivity. This results in a dipole-like artifact in the AF LSSE images from the conventional Seebeck effect, with positive voltage on one side of the boundary and negative voltage on the other. T-domain boundaries are also visible in atomic force microscopy (AFM) as local peaks or valleys, offering a method to check our AF LSSE image interpretation. In Fig.~\ref{AFM comparison}, we compare AF LSSE images with AFM images in both MgO/Pt/NiO bilayer and Pt/NiO/Pt trilayer samples. 

We present raw AF LSSE and AFM images in Fig.~\ref{AFM comparison}(a) and (b), and highlight T-domain walls in (c) and (d). In the bilayer, all major T-domain walls seen in the AF LSSE image are also present in the AFM image. In the trilayer, however, some apparent T-domain lines in the AFM images do not appear in the AF LSSE image.  Because we expect the AFM image to show T-domain walls at both the top and bottom Pt/NiO interfaces, domain walls that do not show in the AF LSSE image indicates that the AF LSSE signal in the trilayers is more sensitive to a single interface. This interpretation agrees with separate imaging of MgO/NiO/Pt and MgO/Pt/NiO bilayers in section S6 that suggests stronger contribution to the AF LSSE signal from the bottom interface than the top.

\subsection{Control measurements on non-magnetic P\MakeLowercase{t}/M\MakeLowercase{g}O}

\begin{figure}[!htb]
\centering
\includegraphics[scale=0.40]{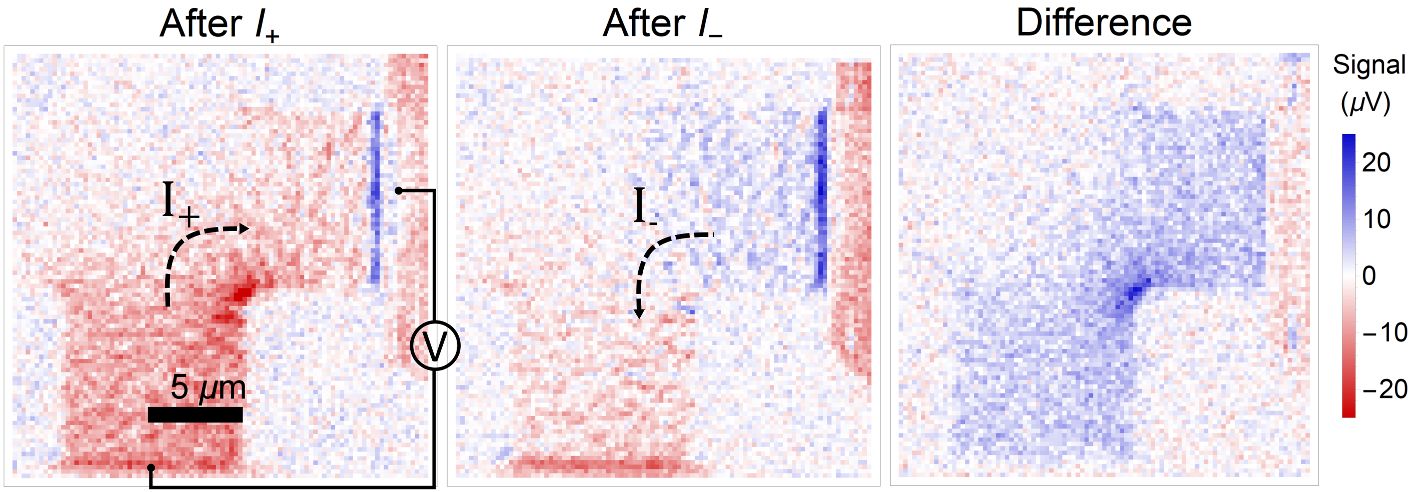}
\caption{Imaging 10 $\mu$m-wide Hall crosses of non-magnetic 10 nm Pt/20 nm MgO to check for non-magnetic artifacts in AF LSSE images. We make voltage contact to the bottom and right branch in an L-shape to apply current along 45$^\circ$ diagonals in the center. Non-magnetic contrast may be due to in-plane Seebeck voltages from local resistance fluctuations. We observe repeatable changes in contrast after applying $4 \times 10^7 \mathrm{A}/\mathrm{cm}^2$ current density. Although the origin of the change in contrast is unknown, it is nearly uniform over the sample and does not exhibit domain flopping or domain motion. }
\label{MGO comparison}
\end{figure}

We perform control images on non-magnetic 10 nm Pt/20 nm MgO, sputtered on sapphire, to check for non-magnetic artifacts in our signal. For example, inhomogeneities in film thickness could cause local inhomogeneities in resistivity, which could lead to voltages from conventional in-plane Seebeck effects. In addition, high local current densities can create sample defects which could appear like switching in the AF LSSE images. We image 10 $\mu$m-wide Hall crosses of Pt/MgO at the same $3.4 \ \mathrm{mJ}/\mathrm{cm}^2$ laser fluence we use for Pt/NiO/Pt, after applying similar current densities. Results are shown in Fig.~\ref{MGO comparison}. Note we make electrical contact to the bottom and right branches in an L-shape instead of the top and bottom contacts as in the Pt/NiO/Pt samples. This is done so that we can apply current along 45$^\circ$ diagonals with only two contacts.  

\begin{figure}[!htb]
\centering
\includegraphics[scale=0.40]{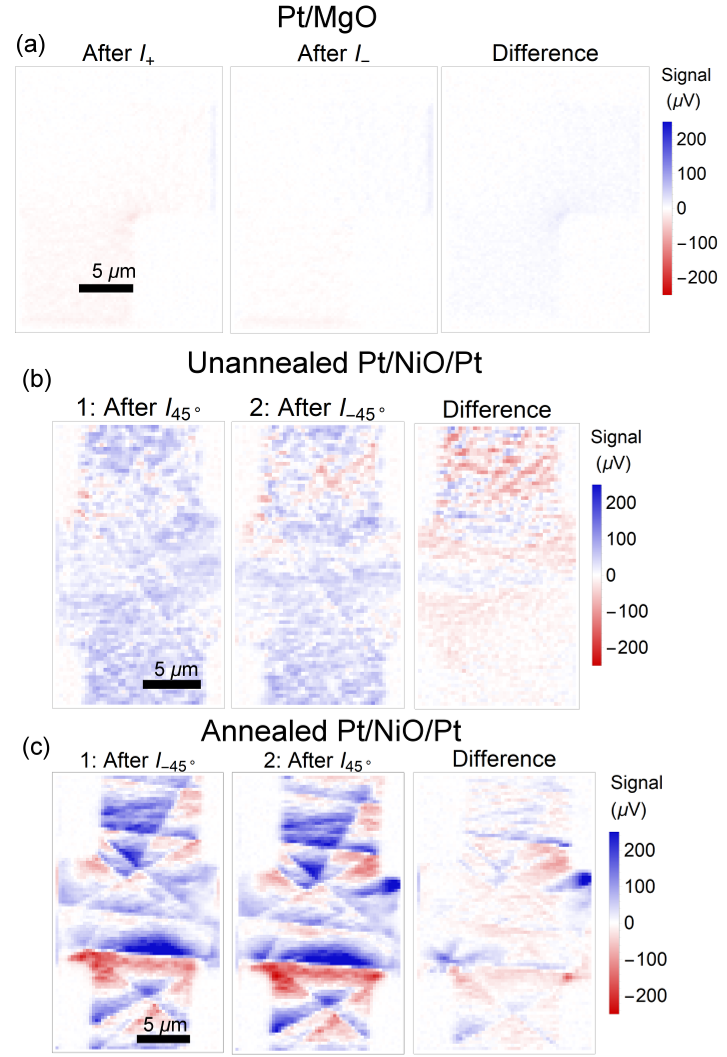}
\caption{Comparing the magnitude of (a) non-magnetic signal from Pt/MgO to AF LSSE signal from (b) unannealed and (c) annealed Pt/NiO/Pt  by plotting them side-by-side on the same color scale. At the same fluence and current density, and normalizing by sample resistance, the magnitude of non-magnetic signal from Pt/MgO is a factor of 5-20 less than the magnitude of AF LSSE signal from Pt/NiO/Pt.}
\label{MgO with NiO}
\end{figure}

We observe submicron contrast which may represent local resistance fluctuations from surface roughness in the Pt layer, which we expect to be more prominent in the polycrystalline Pt/MgO samples than in the epitaxial Pt/NiO/Pt samples. Taking the difference after applying $I_+$ and $I_-$, we observe nearly spatially uniform changes in contrast, which are repeatable. Their origin is unknown: we speculate that they may be due to current-induced motion of sample defects. However, non-magnetic changes in contrast do not resemble local domain flopping or large domain motion. In addition, the overall magnitude of the non-magnetic signal is 5-10 times smaller than the AF LSSE signals from Pt/NiO/Pt, as we show in Fig.~\ref{MgO with NiO} by plotting images of Pt/MgO, unannealed Pt/NiO/Pt, and annealed Pt/NiO/Pt all on the same color scale. We conclude that spurious non-magnetic signal does not contribute significantly to the AF LSSE images of NiO. 

\section{Testing for uncompensated FM moments}

We systematically check for uncompensated moments in Pt/NiO/Pt samples that are distinct from the top and bottom uncompensated AF monolayers. These uncompensated moments would contribute to the $V_{AF~LSSE}$ signal through a ferromagnetic spin Seebeck effect. Possible sources include bulk uncompensated moments in the NiO \cite{SchullerJMMM}, interfacial uncompensated moments separate from the $\lbrace 111\rbrace$ interfacial uncompensated AF monolayers \cite{HillebrechtPRL}, canted moments at the Pt/NiO interfaces from symmetry breaking, and proximity-induced magnetization in the Pt. We first take AF LSSE images of an unannealed Pt/NiO/Pt sample at $\pm$2.5 kG, the largest field we can apply in our setup, as shown in Fig. \ref{images vs field} below. 

\begin{figure}[!htb]
\centering
\includegraphics[scale=0.40]{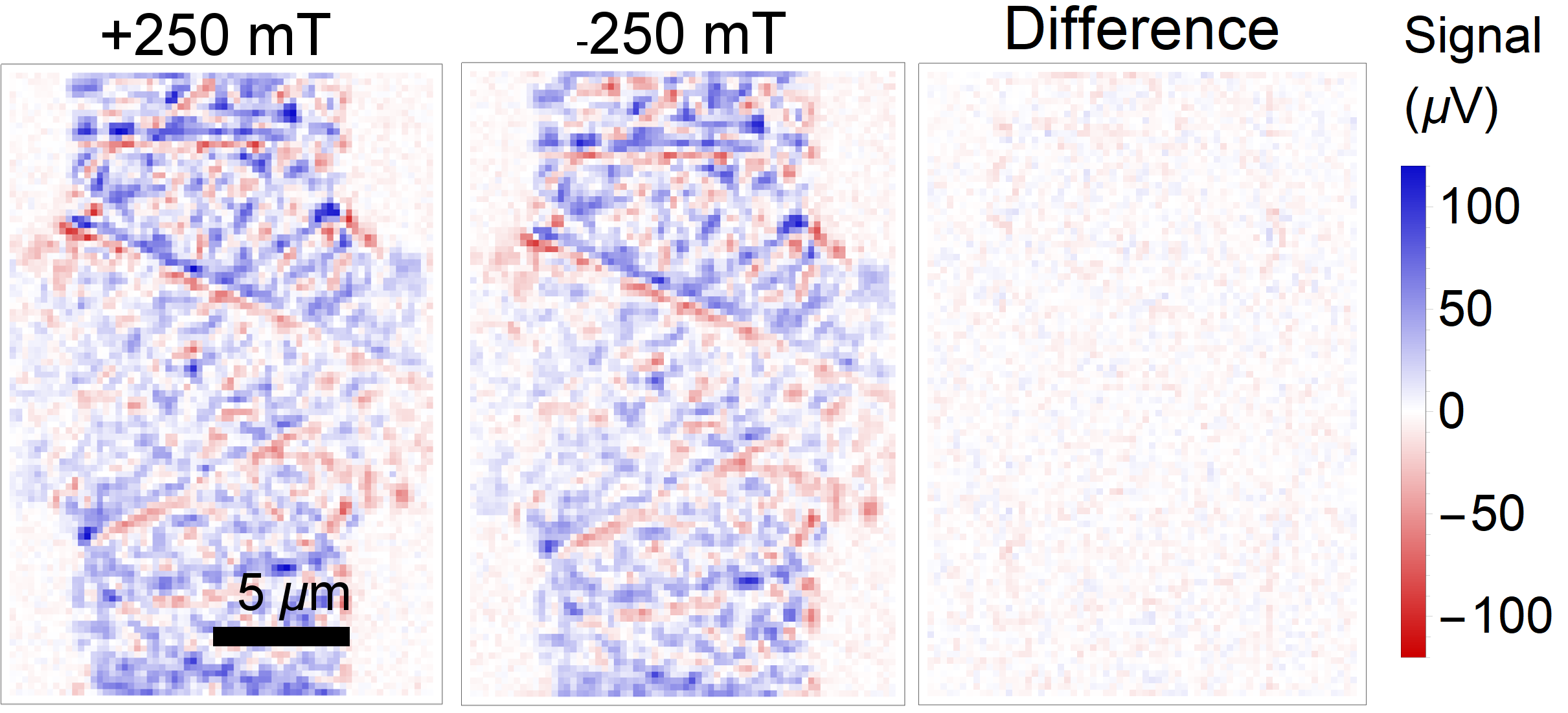}
\caption{AF LSSE images of Pt/NiO/Pt at $\pm$2.5 kG magnetic field. We observe no change in contrast within our sensitivity, which is consistent with an antiferromagnetic AF LSSE origin since the spin-flop field for NiO is 7 T and the threshold field for domain wall motion is 1.5 T.}
\label{images vs field}
\end{figure} 

The AF LSSE signal is nearly pixel-for-pixel identical at $\pm$2.5 kG, which we expect since the spin-flop field in NiO is near 7 T \cite{MachadoPRB} and the threshold field for domain motion is 1.5 T \cite{SaitoJPhysC}. 

\begin{figure}[!htb]
\centering
\includegraphics[scale=0.8]{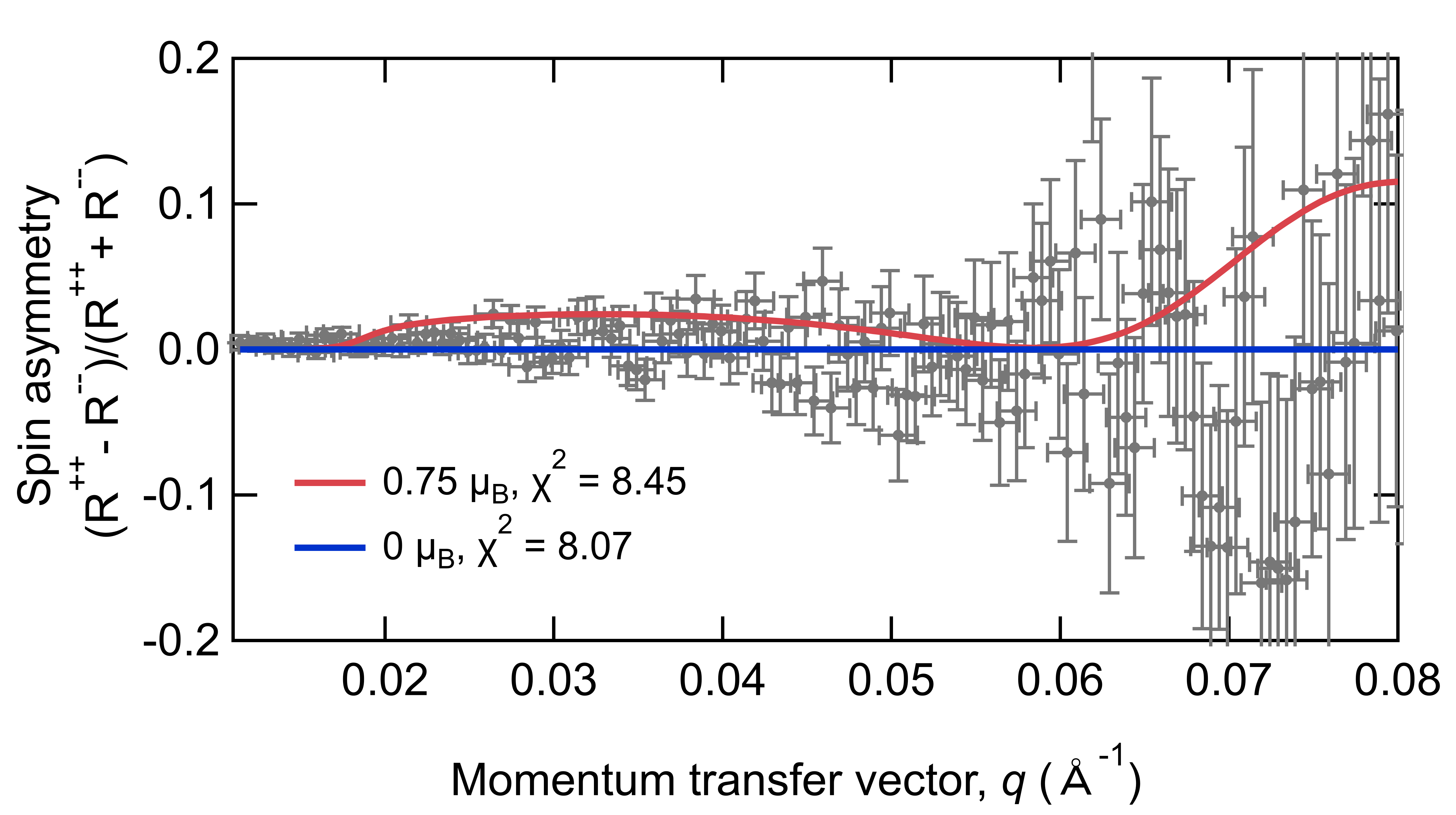}
\caption{The spin asymmetry plot from polarized neutron reflectometry on an unpatterned Pt/NiO/Pt film, performed at 295 K and 0.7 T applied field. No net moment is observed. From the noise level, we estimate an upper bound on the magnetization from one monolayer of 0.75 $\mu_B$/Ni atom.}
\label{PNR}
\end{figure}

We perform polarized neutron reflectometry (PNR) on an unpatterned Pt/NiO/Pt film at room temperature to measure any overall magnetic moment in the film stack at 0.7 T applied field. The spin asymmetry plot is shown in Fig.\ref{PNR}. Within our sensitivity we measure no spin asymmetry and thus no net moment. We place an upper bound on the magnetization that could be present by modeling the expected spin asymmetry from a single polarized monolayer, which we choose in order to simulate uncompensated interfacial moments. We obtain a maximum magnetization of 0.75 $\mu_B$/Ni, ruling out a fully magnetized (1.9 $\mu_B$/Ni) monolayer.

\begin{figure}[!htb]
\centering
\includegraphics[scale=0.35]{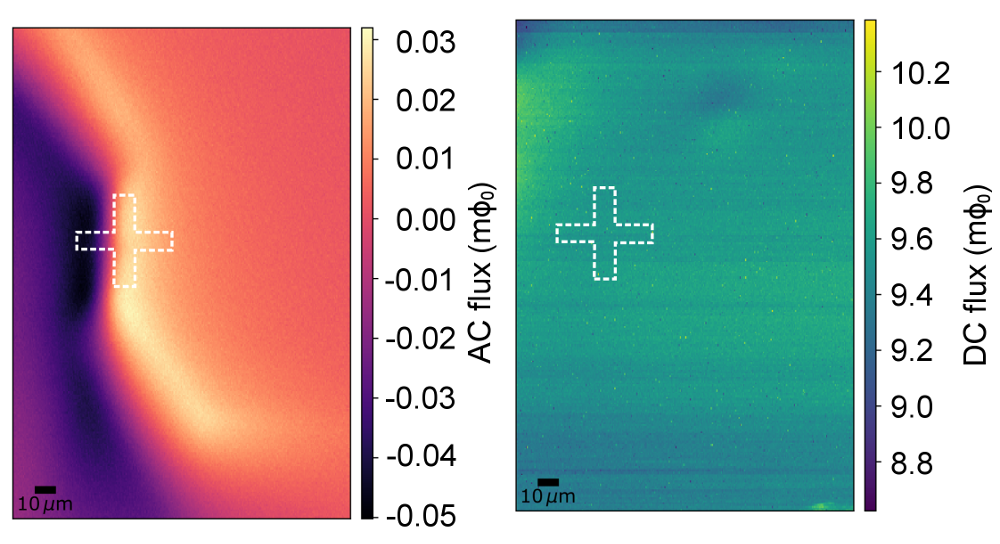}
\caption{Scanning SQUID images of an annealed 10 $\mu$m-wide Pt/NiO/Pt Hall cross at 7~K, with $\sim1 \ \mu$m resolution. (a) Magnetic flux generated by running 1 mA current through the sample. (b) Flux after turning off the current. The sample edges are outlined in white. From the lack of contrast observed, we estimate an upper bound of $8 \times 10^{-4} \  \mu_B$/Ni local moment from a magnetized surface monolayer.}
\label{SQUID}
\end{figure}

Uncompensated moments could still be present if they are pinned by the N\'eel order and the N\'eel orientation averages to zero on the scale of microns to tens of microns. In Figure~\ref{SQUID} we search for local moments by performing scanning SQUID microscopy, which directly images magnetic flux with $\sim1 \ \mu$m resolution, on an annealed 10 $\mu$m-wide Pt/NiO/Pt Hall cross at a temperature of 7 K.

In Fig.~\ref{SQUID}(a) we locate the sample by imaging the flux while applying DC current through the vertical branch, and then in Fig. S11(b) we image the same region again with no current applied. The sample edges are outlined in white. Out-of-plane moments would produce signal in the width of the channel, while in-plane moments would produce signal at the sample edges. However, Fig.~\ref{SQUID}(b) shows no contrast within sensitivity, so we do not measure any moment. By modeling the expected response from a magnetized surface monolayer, similarly to the PNR results, we calculate an upper bound of $8 \times 10^{-4} \ \mu_B$/Ni. From this value we calculate a maximum bulk magnetization of 110 A/m, three orders of magnitude less than the 140 kA/m bulk magnetization of YIG at room temperature. The combination of insensitivity of the AF LSSE signal to magnetic field with null results from PNR and scanning SQUID leads us to conclude that the AF LSSE image signal originates from an antiferromagnetic spin Seebeck effect in NiO, rather than a ferromagnetic SSE from uncompensated moments. 

\section{Averaging process for AF LSSE images}

\begin{figure}[!htb]
\centering
\includegraphics[scale=0.35]{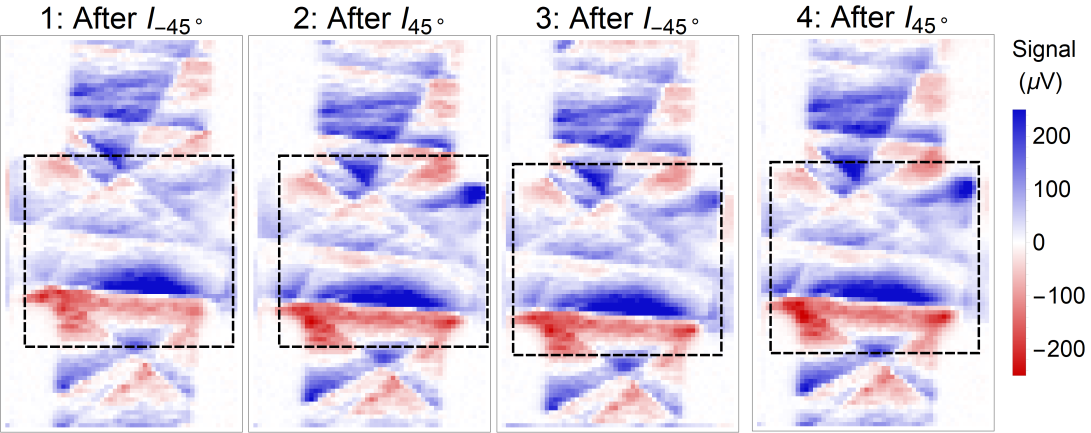}
\caption{The AF LSSE images from Figure 3 of the main text. We obtain the average $V_{AF~LSSE}$ values in Fig. 3(c) and Fig. 3(e) by averaging all the pixels in each image within a 12~$\mu$m~$\times$~12~$\mu$m square region centered on the cross center, represented by the dashed lines.}
\label{averaging}
\end{figure}

In the main text, we compare images of the N\'eel orientation with the electrical readout by averaging the pixels in the AF LSSE images in and near the center of the 10 $\mu$m-wide Hall cross and comparing to $R_H$. As shown in Figure~\ref{averaging}, we average pixels within a $12$ $\mu$m-wide square centered on the cross center to ensure that contributions from the corners are incorporated. The correct dimensions that should be used for the averaging window are not obvious, since current flow is non-uniform within the cross. Therefore, the error bars on the average $V_{AF~LSSE}$ in Fig.~3(e)in the main text are determined by calculating the change in the average $V_{AF~LSSE}$ after varying the dimensions of the averaging window by 2 pixels in both the x and y-directions. In principle the pixels within the averaging window should also be weighted by the spatially-varying current density, but as a first approximation we average all pixels equally.   

\section{Effects of roughness and two P\MakeLowercase{t}/N\MakeLowercase{i}O interfaces on the AF LSSE signal} 

\subsection{Effects of interfacial roughness at the P\lowercase{t}/N\lowercase{i}O interface}

\begin{figure}[!htb]
\centering
\includegraphics[scale=0.09]{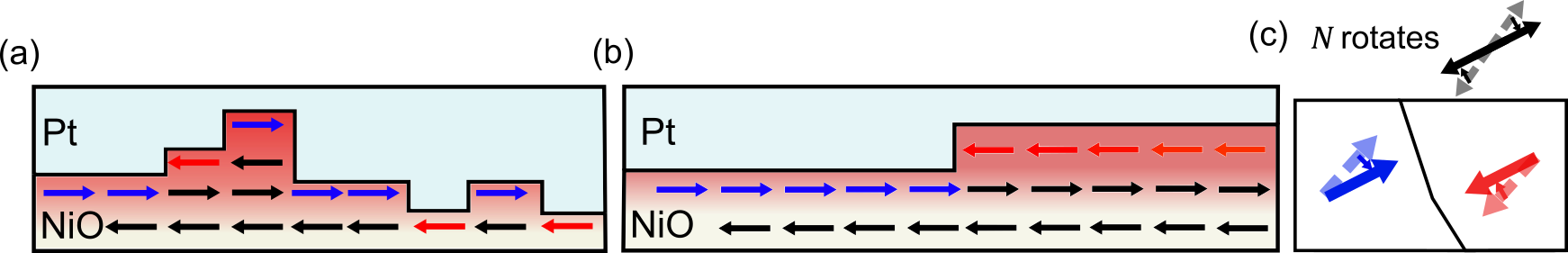}
\caption{Potential effects of surface roughness on the AF LSSE signal. Region 1 shows the ideal, atomically flat interface. Region 2 represents roughness in the form of random atomic steps, which decreases the overall signal but does not change its interpretation. Region 3 shows a flat interface with an average height one monolayer different than Region 1, resulting in the opposite sign in the AF LSSE signal within the same S-domain.}
\label{roughness diagram}
\end{figure}

Here we discuss the interfacial AF LSSE in the context of real samples, with surface and interface roughness. Surface defects could cause artifacts in the AF LSSE images, for example, from resistance variations in the Pt layer. These artifacts originate from the charge Seebeck effect and are non-magnetic. We can distinguish them from AF LSSE signal by taking difference images.  Local height variations in the NiO \cite{CharilaouJAP} could have a greater effect on the overall AF LSSE signal because an atomic step of one monolayer could cause $\bm{J}_s$ from the AF LSSE to locally reverse direction, as shown in Fig.~\ref{roughness diagram}. We distinguish two possible effects of random height variation depending on its lateral length scale. If the lateral length scale of surface roughness is much less than the 650 nm laser spot size, the laser averages over the roughness, as shown in Fig.~\ref{roughness diagram}(a). From AFM scans, we estimate the surface height variation to be about 1 nm, or 2-3 unit cells, and the lateral length scale of the height variation to be tens of nanometers. Although the sign of the averaged signal depends on the exact distribution of the height steps, the presence of atomic steps at the nanometer scale does not fundamentally alter the interpretation of the AF LSSE voltage as reporting $N_x$ with a consistent sign inside S-domains. 

If the lateral scale of height variation is comparable to or greater than the laser spot size -- in other words, if the average thickness varies from pixel to pixel, as illustrated in Fig.~\ref{roughness diagram}(b), we would expect the AF LSSE voltage to contain contributions from both the spin structure and local thickness variations, which would manifest as nonuniformity and sign changes in the AF LSSE signal within a single S-domain. 

We argue that local thickness variations does not significantly contribute to the AF LSSE signal. First, the annealed samples in Fig.~3 in the main text, as well as those shown in Fig.~\ref{temp dependence} and \ref{AFM comparison}, contain S-domains that are 3-10 $\mu$m wide. From AFM images, we know that the sample thickness is not uniform to within one monolayer over a distance of several $\mu$m. Therefore, if thickness variations made a significant contribution to our signal, we would expect to observe signal variation, including sign changes, inside the S-domain. Instead, we find the contrast is almost uniform within the S-domains.

Second, AF LSSE contrast from thickness variations would exhibit characteristic switching patterns that we do not observe. In Fig.~\ref{roughness diagram}(c) we illustrate how a uniform S-domain with a thickness step of one monolayer (the most extreme case) would respond under domain rotation. Upon first imaging, we would obtain a sign reversal as the probe crosses the step. After imaging domain rotation from spin torque, we expect $\Delta V_{AF LSSE}$ in the two regions to be anticorrelated -- if the blue region becomes bluer, the red region should become redder, and vice versa. Therefore, if signal contrast were dominated by interfacial steps, the difference image should inherit the pixel-to-pixel nonuniformity of the AF LSSE images. In the AF switching images in Fig.~4 of the main text, we see the opposite effect: the difference images are more uniform than the AF LSSE images themselves. Within the region labeled \textit{domain rotation}, blue domains become more blue and red domains less red, corresponding to a uniform increase in $N_x$. In addition, thickness variations would not move under applied current, therefore the domain wall motion and domain expansion in Fig.~4 of the main text is not due to surface roughness. 

Third, although we expect contrast due to thickness variations not to be affected by heat treatment up to the 200 $^\circ$C annealing temperature, we find that S-domains in annealed samples are more uniform and an order of magnitude larger than in unannealed samples. Based on these reasons, we conclude that in our samples the AF LSSE images represent spin contrast and not surface roughness. Further studies, perhaps comparing different growth methods, are necessary to elucidate the role of surface roughness in interfacial AF LSSE.
 
\subsection{Effects of two Pt/NiO interfaces on the AF LSSE signal}

\begin{figure}[htb]
\centering
\includegraphics[scale=0.38]{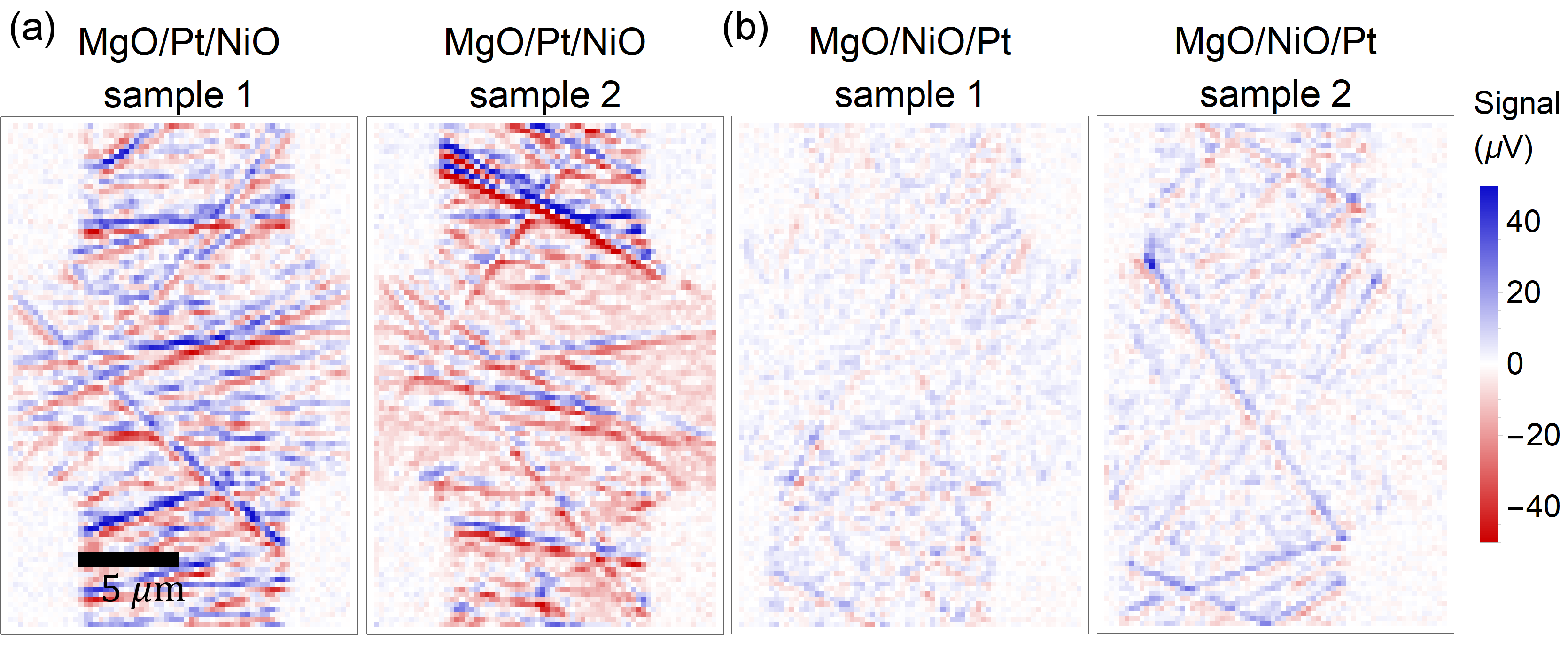}
\caption{Control LSSE images of (a) MgO/4 nm Pt/10 nm NiO and (b) MgO/10 nm NiO/4 nm Pt, to separately image spin Seebeck signal from the bottom and top Pt/NiO interfaces, respectively. Weaker contrast from the top interface in (b) compared to the bottom in (a) may reflect differences in interface quality.}
\label{single layer NiO}
\end{figure}

 Finite-element simulations of laser heating in Pt/NiO/Pt trilayers in Fig.~\ref{bilayer simulations} and \ref{trilayer simulations} show that the thermal depth profile is dominated by temperature discontinuities at the Pt/NiO interfaces. The temperature drop across each interface is nearly equal, suggesting that both interfaces can contribute to the AF LSSE signal in trilayers. This is a potential difficulty if the sign of the uncompensated interfacial spins is opposite for the top and bottom interface, or if the spins at the top and bottom interfaces represent two different S-domains. Therefore, we measure the contributions of the two interfaces separately by imaging control samples of unannealed MgO/4 nm Pt/10 nm NiO and MgO/10 nm NiO/4 nm Pt, patterned into 10 $\mu$m $\times$ 50 $\mu$m Hall crosses, using the same $3.4 \ \mathrm{mJ}/\mathrm{cm}^2$ fluence we use in the main text. Results are shown in Fig.~\ref{single layer NiO}. 

T-domains are visible in all images and S-domain contrast is visible at least in sample 2 in Fig.~\ref{single layer NiO}(a). T-domain walls are more numerous and prominent in images of a single interface, seen especially in Fig.~S14(a), than in the Pt/NiO/Pt images. In addition, the LSSE signal from the NiO/Pt interface in (b) is a factor of 5 weaker than the signal from the Pt/NiO interface in (a). This result is not suprising, because the AF LSSE voltage we measure depends on the spin mixing conductance at the Pt/NiO interface, which in turn depends on surface roughness, and we expect the bottom interface to be less rough than the top. 

\begin{figure}[!htb]
\centering
\includegraphics[scale=0.11]{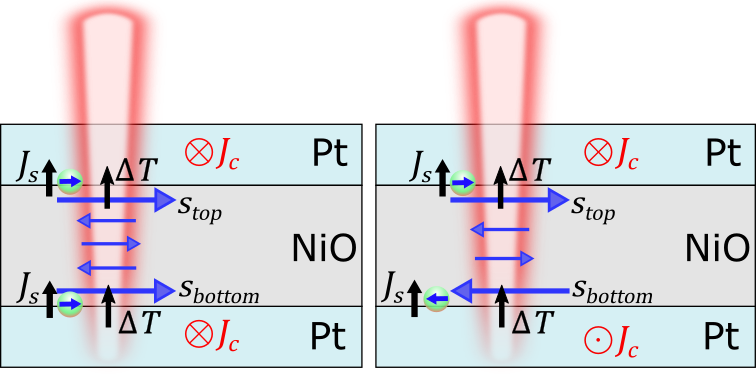}
\caption{Schematic showing the interfacial spin Seebeck voltage produced from two Pt/NiO interfaces. $\mathbf{J}_c$ adds together when the interfacial uncompensated spins are parallel (a) and cancels out when the spins are antiparallel (b).}
\label{two interface schematic}
\end{figure}

Even though the single-interface images suggest that the bottom NiO/Pt interface contributes more to the net AF LSSE signal than the top, we consider the effects of different interface terminations, assuming a continuous S-domain in thickness, in Fig.~\ref{two interface schematic}. Simulations in Fig.~\ref{trilayer simulations} show that the sign of the interfacial thermal gradient $\Delta T$ is the same at both interfaces. The sign of the spin current $\bm{J}_s$ must follow that of the thermal gradient in linear response. Using $\bm{J}_c \propto \bm{J}_s \times \bm{\sigma}$, we show in Fig.~\ref{two interface schematic} that $\bm{J}_c$ and therefore the AF LSSE voltage from each interface adds together when the top and bottom spins are parallel (a) and cancels out when they are antiparallel (b).

This result would initially seem to hinder imaging of switching in trilayers, because only regions with antiparallel top and bottom spins contribute to switching according to the argument of Refs. \cite{MoriyamaArxiv} and \cite{MoriyamaPRL}. However, we argue AF LSSE imaging should still show switching, as follows: although we expect that spatial variation between parallel and antiparallel spins occurs on the nanometer scale, we also expect that the exchange coupling between spins causes domain rotation on the scale of hundreds of nanometers. Therefore the parallel regions that contribute to the AF LSSE signal are dragged along by the antiparallel regions that contribute to switching. The 650 nm laser spot size averages out surface roughness to yield a net signal that is proportional to $N_x$.

\section{Domain wall creep in in a P\lowercase{t}/N\lowercase{i}O bilayer}

\begin{figure}[!htb]
\centering
\includegraphics[scale=0.9]{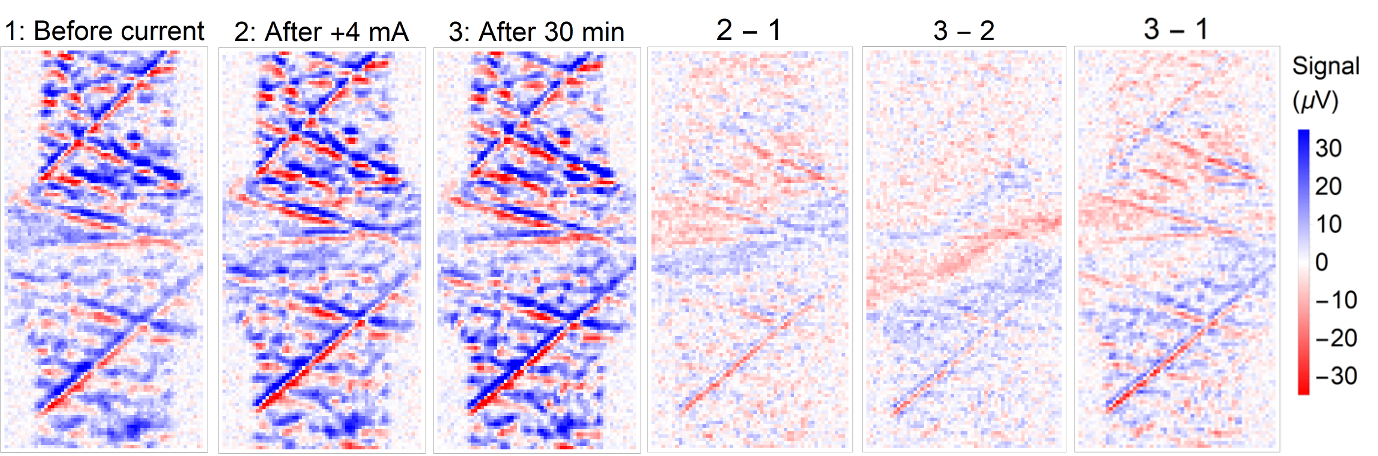}
\caption{AF LSSE images of the Pt/NiO bilayer from Fig.~4 in the main text before applying current, after applying 4 mA ($1\times 10^7~\mathrm{A}/\mathrm{cm}^2$), and after 30 minutes without stimulus. Initial domain rotation is followed by domain wall motion.}
\label{domain wall motion}
\end{figure}

Fig.~4 in the main text shows domain rotation and domain wall motion in a MgO/Pt/NiO bilayer in response to current densities between $5.0\times 10^7~\mathrm{A}/\mathrm{cm}^2$ and $1.1\times10^8~\mathrm{A}/\mathrm{cm}^2$. In Fig.~\ref{domain wall motion} we image after applying $1.0\times 10^7 ~\mathrm{A}/\mathrm{cm}^2$, image again after 30 minutes, and take sequential and cumulative image differences.

We observe apparent domain rotation immediately after switching in $\mathit{2 - 1}$. $\mathit{3-2}$ shows domain wall motion after 30 minutes, even with no stimulus (also shown later in Section S8). The cumulative difference in $\mathit{3 - 1}$ resembles the differences in Fig.~4(b) in the main text, but fainter. We speculate that after the spin torque rotates the S-domains out of equilibrium, magnetoelastic stresses exert forces on the domain wall which result in subthreshold domain wall creep. As discussed in the main text, the switched states after +30 mA and +42 mA are stable in time, meaning they do not relax back, but the domain wall motion after +42 mA requires less current to reverse (at most -20 mA), which suggests that the switched state is metastable.

\section{AF LSSE images of switching in unannealed P\lowercase{t}/N\lowercase{i}O/P\lowercase{t} at $I_{\pm 45^\circ}$}

\begin{figure}
\centering
\includegraphics[scale=0.63]{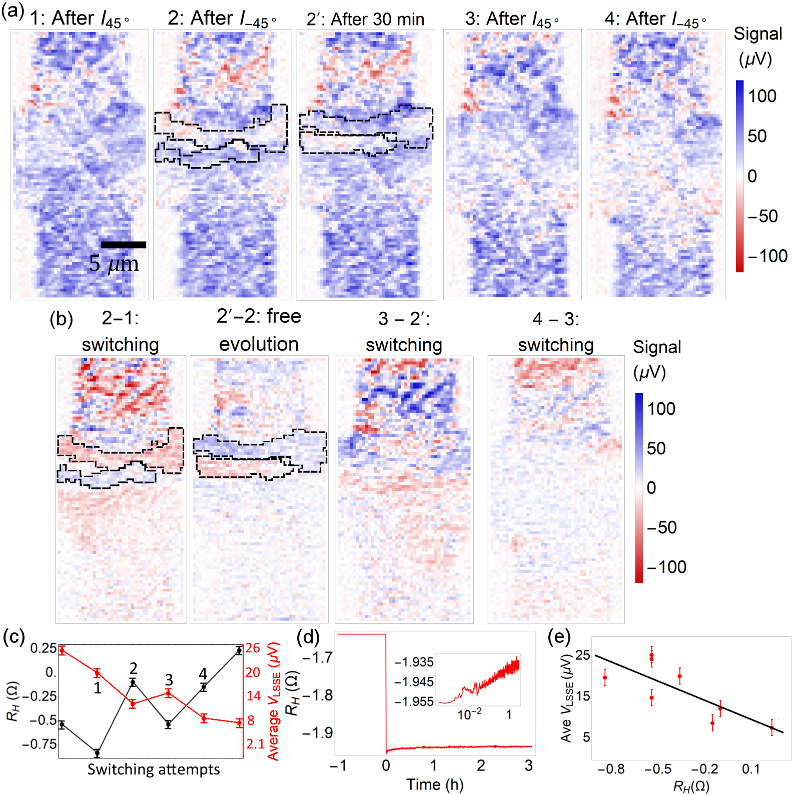}
\caption{Switching at $I_{\pm 45^\circ}$ in an unannealed  trilayer, where the switching efficiency measured with $\Delta R_H/R_H$ is increased by a factor of 5 (0.02$\%$ to $0.1\%$) from the annealed sample in Fig.~3 in the main text. (a) AF LSSE images while toggling between $I_{\pm 45^\circ}$ as in Fig.~3 of the main text. In addition, we image twice after applying current in $\mathit{2}$ and $\mathit{2}'$ to resolve domain wall creep. (b) Sequential differences between the images in (a). In addition to domain rotation above the cross center, we observe stripes of alternating sign that correspond to a combination of domain rotation and domain wall motion, highlighted in dashed line. $\mathit{3} - \mathit{2}$, taken after 30 minutes, shows that after switching the domains relax in the opposite direction in which they initially rotated or moved. Domain wall motion decreases with repeated toggling in a training-like effect. (c) $R_H$ and $\langle V_{AF~LSSE} \rangle$ while toggling between $I_{45^\circ}$ and $I_{-45^\circ}$. Values of $R_H$ and $\langle V_{AF~LSSE} \rangle$ corresponding to AF LSSE images are labeled. Decreased domain motion corresponds to decreased $\Delta R_H$. (d) $R_H$ vs time after switching in a similar sample. The inset shows that $R_H$ approximately exponentially relaxes over 2-3 hours opposite to the direction of initial switching, consistent with the domain wall creep that we image. (e) $\langle V_{AF~LSSE} \rangle$ vs  $R_H$. Although the switching is less uniform than in Fig.~3 in the main text, we still find a correlation.} 
\label{multiple switching}
\end{figure}

In Fig.~3 of the main text, we study switching of an annealed Pt/NiO/Pt trilayer at $I_{\pm 45^\circ}$. In Fig.~\ref{multiple switching}, we perform the same switching and imaging procedure on an unannealed Pt/NiO/Pt trilayer, where the SMR response is 5 times larger ($\Delta R_H/R_H = 0.1\%$ compared to $0.02 \%$.)  We present AF LSSE images after applying current at $\pm 45 ^\circ$ in Fig.~S17(a), and additionally image after waiting 30 minutes without applying current in image $\mathit{2}'$ to resolve domain creep. We take AF LSSE image differences in Fig.~\ref{multiple switching}(b) and compare to concurrently taken SMR measurements in Fig.~\ref{multiple switching}(c). 

We find switching patterns consistent with the domain rotation and domain wall motion identified in Fig.~4 in the main text, although the signal-to-noise in this sample is lower. In the difference images in Fig.~\ref{multiple switching}(b), we observe relatively large-scale contrast that is more uniform than the AF LSSE images themselves, indicating uniform S-domain rotation from the out-of-plane component of the spin torque as described in the main text. The regions of rotation are highlighted in dashed line in both Fig.~\ref{multiple switching}(a) and (b).  Slight changes in contrast in the AF LSSE signal near the center of the cross show that the S-domains rotate by acute angles, similar to Fig.~3 in the main text.

 Inspection of $\mathit{2}'-\mathit{2}$, which represents the difference between the N\'eel order just after current is applied and after 30 minutes without stimulus, reveals that the S-domains continue to evolve after the current is removed. In particular, we see that they relax opposite to the direction in which they were pushed by the current. We see a corresponding time variation in the SMR in a similar sample: Fig.~\ref{multiple switching}(d) shows $R_H(t)$ after switching at the same current density. After switching, $R_H$ relaxes approximately exponentially in the direction opposite to its initial switch, as predicted by models of subthreshold magnetoelastic domain creep, over a 3-hour period. 
 
Fig.~\ref{multiple switching}(b) also reveals that domain rotation and domain wall motion is diminished after the third successive switching attempt at the same current density. By measuring $R_H$ after each image, we find that the switching efficiency decreases by a factor of 2, and continues to decrease until $\Delta R_H$ stabilizes at 60 m$\Omega$, 13 times smaller than the initial 750 m$\Omega$ change. Decreased switching efficiency after repeated switching is consistent across all the samples we measure. This behavior may be analogous to the training effect in exchange-biased antiferromagnet/ferromagnet bilayers, where the magnitude of exchange bias and coercivity decrease after repeated reversals. 

In Fig.~\ref{multiple switching}(c), we track sequential averaged AF LSSE signal and the concurrently measured $R_H$, and in Fig.~\ref{multiple switching}(e) we plot one versus the other, similarly to Fig.~3(e) and Fig.~3(f) in the main text. As discussed in the main text, in general we cannot directly compare $\langle V_{AF~LSSE}\rangle$ to $R_H$, since they have different symmetries. We can make a correspondence in Fig.~3 of the main text, since most of the change in contrast is localized to a spot on the corner where $N_x$ is nearly saturated. Although the changes in contrast in Fig.~\ref{multiple switching} are less spatially uniform than in Fig.~3, most of the contrast is still positive $N_x$ of similar hue. Therefore, $\langle V_{AF~LSSE} \rangle$ still roughly tracks $R_H$ in Fig.~\ref{multiple switching}(c), and plotting one versus the other in Fig.~\ref{multiple switching}(e) shows a correlation, although not as strong as the correlation in Fig.~3. of the main text.   

Comparing the unannealed sample with the annealed sample in Fig. 3 of the main text, we find that the S-domains in the unannealed sample are submicron in size, while the domains in the annealed sample are 2-10 $\mu$m in size. The unannealed sample exhibits more prominent domain wall motion than the annealed sample, consistent with the factor-of-5-larger SMR response, which suggests that the larger and uniform S-domains in the annealed sample require a higher threshold current for domain wall motion. This opens up the possibility of increasing switching efficiency by controlling the S-domain size, which will require further imaging studies.
\section{Characterization of spin-torque switching}

\subsection{Experimental procedure for spin-torque switching and SMR reading}

We use a Keithley 2400 sourcemeter for both spin-torque writing and electrical reading in all samples. The switching in Fig. 3 of the main text is done with DC current, applied for 5 s using a current density of $3.1\times10^7~\mathrm{A}/\mathrm{cm}^2$. Note that while applying current at $\pm 45^\circ$ requires four electrical contacts, AF LSSE microscopy only works with two contacts: stray capacitance between four contacts at high frequencies causes signal leakage, making the images difficult to interpret. Therefore, we wire bond the two AF LSSE voltage contacts, and we perform spin-torque switching and SMR measurement using a probe station to make temporary contacts. In Fig. 4, where we apply writing current along the $V_{AF~ LSSE}$ voltage contacts, we do not have this difficulty. In the sample in Fig.~4, we apply writing current using a series of 10 2 ms-wide DC pulses.  

\subsection{SMR while toggling $I_{\pm 45^\circ}$}

\begin{figure}
\centering
\includegraphics[scale=0.50]{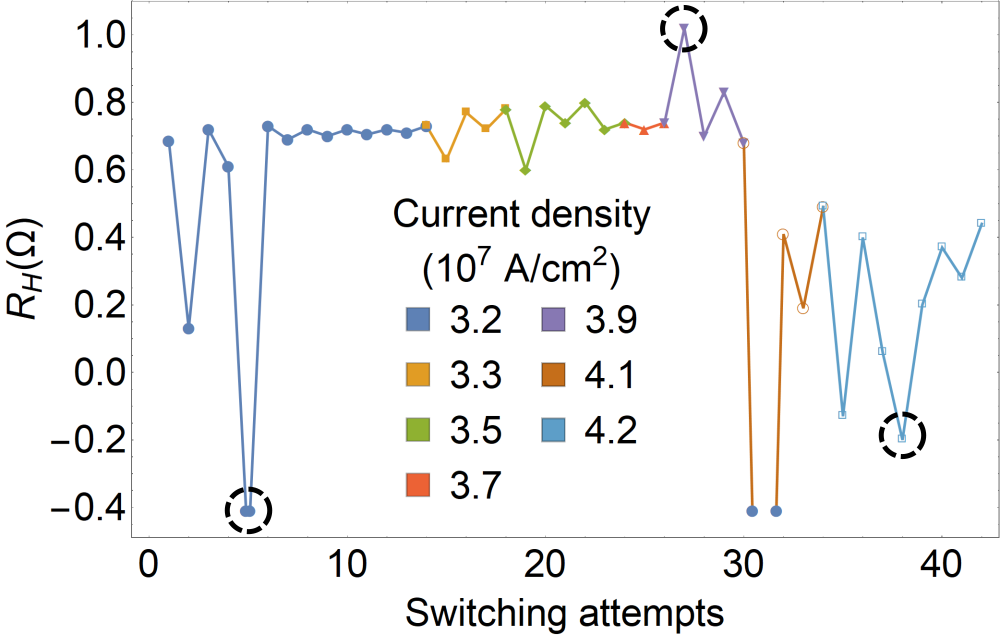}
\caption{$R_H$ in a 15 $\mu$m-wide Pt/NiO/Pt cross as a function of writing current density, alternating between $I_{45^\circ}$ and $I_{-45^\circ}$ at each point. An initial $\Delta R_H$ of 100s of $\mathrm{m}\Omega$ is consistently followed by $\sim 50 \mathrm{m}\Omega$ changes, which may be correlated with the decrease in domain wall motion that we observe in LSSE images. At 3.2, 3.9, and 4.2 $\times 10^7 \mathrm{A}/\mathrm{cm}^2$, highlighted in the plot, the polarity of the switching seems to reverse, which may be due to different threshold current densities for switching in different spatial regions of the sample.}
\label{electrical switching}
\end{figure}

To characterize both reproducibility of spin-torque switching at a given writing current density and the magnitude of switching as a function of current density, we measure $R_H$ in an unannealed 15 $\mu$m-wide Hall cross after toggling between $45^\circ$ and $-45^\circ$ writing current directions as described in the main text. Results are shown in Fig.~\ref{electrical switching}: each point represents $R_H$ after toggling the current direction.

We find a much larger (100s of m$\Omega$) change in $R_H$ after the first several switching attempts at a given current density than the $\sim$50 m$\Omega$ changes that follow, which is consistent with a decrease in antiferromagnetic domain wall motion. Surprisingly, the polarity of switching seems to reverse, between $\bm{N}$ rotating towards $\bm{J}_W$ and $\bm{N}$ rotating perpendicular to $\bm{J}_W$, at $3.2 \times 10^7 \mathrm{A}/\mathrm{cm}^2$, $3.9 \times 10^7 \mathrm{A}/\mathrm{cm}^2$, and $4.2 \times 10^7 \mathrm{A}/\mathrm{cm}^2$. This may be due to the chiral domain force $\bm{F}_{DW}$ documented by Ref. \cite{BaldratiArxiv} and our work, which can result in both $\bm{N} \parallel \bm{J}_W$ and $\bm{N} \perp \bm{J}_W$, combined with different current thresholds for different regions of the sample. Further studies are necessary to determine long-term switching reproducibility in multidomain samples.  

\section{Finite-element calculations of laser heating}

We perform finite-element calculations of laser heating in Pt/NiO/Pt trilayers using the COMSOL Multiphysics$^{\tiny{\textregistered}}$ software package. We calculate the temperature profile by solving the radially symmetric heat diffusion equation, modeling the laser as a distributed heat source that exponentially decays with thickness according to the skin depths of Pt and NiO.

\begin{figure}
\centering
\includegraphics[scale=0.31]{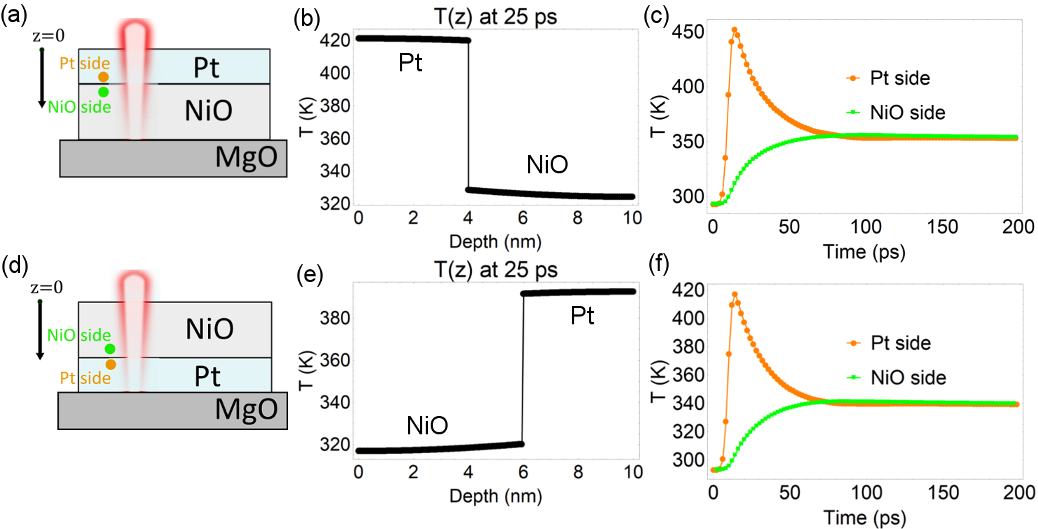}
\caption{(a) Schematic of the NiO/Pt bilayer. (b) Thermal depth profile at 25 ps, near peak heating. Most of the thermal gradient occurs at the Pt/NiO interface. (c) Temperature vs time at two points 0.5 nm above and below the Pt/NiO interface. The Pt layer heats to a maximum of 450 K, while the NiO layer heats to 360 K. (d-f) The same plots for the Pt/NiO bilayer. Again the Pt layer heats more than the NiO, so the sign of the thermal gradient is reversed.}
\label{bilayer simulations}
\end{figure} 

We first separately model heating in MgO/Pt/NiO and MgO/NiO/Pt bilayer samples in Fig.~\ref{bilayer simulations} at the 3.4 $\mathrm{mJ}/\mathrm{cm}^2$ fluence used in the experiment. We plot in Fig.~\ref{bilayer simulations}(b) and (e) the temperature depth profile near peak heating at 25 ps in the Pt/NiO and NiO/Pt samples, respectively. The profiles are dominated by the thermal resistance between Pt and NiO, assuming an interface thermal conductance of 500 $\mathrm{MW}/\mathrm{m}^2 \cdot \mathrm{K}$, which is high even for epitaxial interfaces \cite{CostescuPRB} and thus probably an overestimation. Therefore, in Fig.~\ref{bilayer simulations}(c) and (f) we plot temperature vs time for two points 0.5 nm above and below the interface. Because NiO is nearly transparent at the 785 nm laser wavelength, the Pt layer heats more than the NiO. The sign of the thermal gradient is therefore opposite for NiO/Pt than for Pt/NiO. We estimate peak heating of the NiO layer to be 55 K for Pt on top and 45 K for Pt on the bottom. We estimate about 5 K background heating, because the sample does not completely cool to room temperature after 13 ns, when the next pulse arrives. 

\begin{figure}
\centering
\includegraphics[scale=0.38]{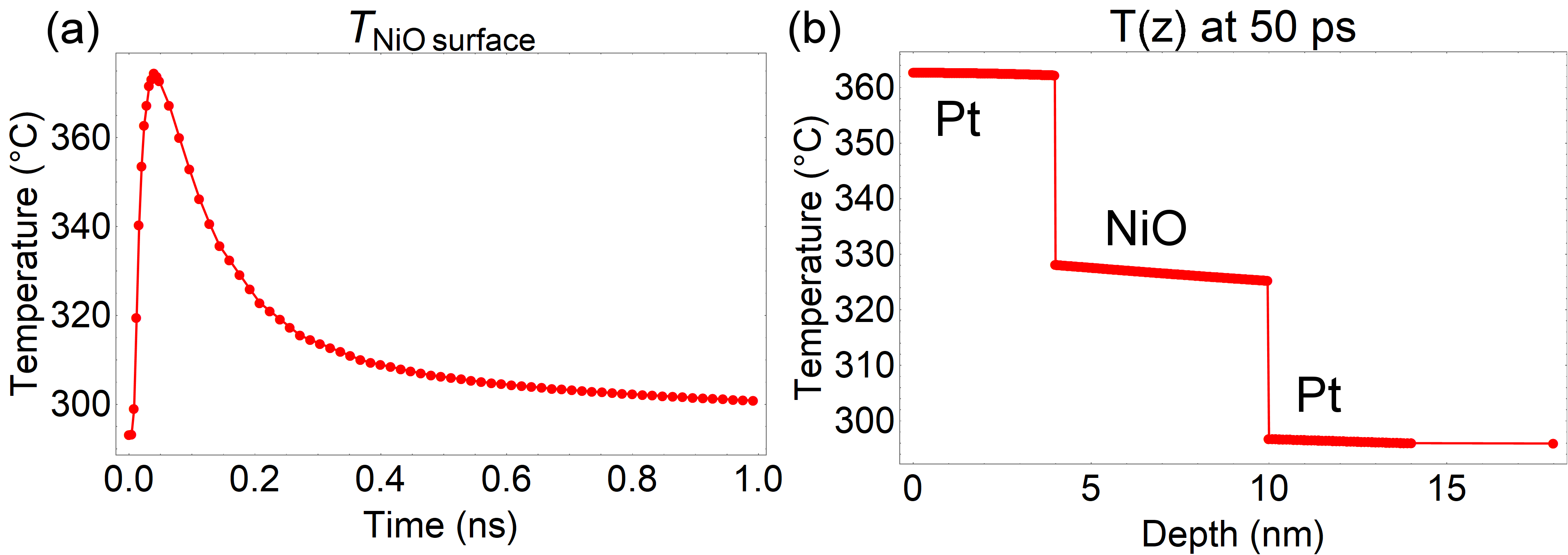}
\caption{Finite-element calculations of laser heating. (a) The temperature at the NiO surface, following a 3 ps laser pulse at 3.4 $\mathrm{mJ}/\mathrm{cm}^2$ fluence. We estimate a peak temperature change of $70 ^\circ$C. (b) Temperature as a function of depth at peak heating, 50 ps after pulse arrival. We assume a thermal conductance of 500 $\mathrm{MW}/\mathrm{m}^2 \cdot \mathrm{K}$ at both Pt/NiO interfaces, which is typical for epitaxial, lattice-matched interfaces \cite{CostescuPRB}. The temperature drop across each interface is about $30 ^\circ$C compared with $2 ^\circ$C drop across the NiO thickness, leading us to conclude the interfacial spin Seebeck effect is more strongly excited than the bulk. }
\label{trilayer simulations}
\end{figure}

We then simulate laser heating of the Pt/NiO/Pt trilayers in Fig.~\ref{trilayer simulations} at the same fluence. In this case the sign of the thermal gradient is the same at both interfaces. In Fig.~\ref{trilayer simulations}(a) we plot the temperature at the NiO surface as a function of time after the heating pulse arrives. We estimate that at 3.4 $\mathrm{mJ}/\mathrm{cm}^2$ fluence, the top surface of the top Pt layer reaches a maximum temperature of 400 K, and the top surface of the NiO reaches a maximum temperature of 370 K.

In Fig.~\ref{trilayer simulations}(b) we plot the temperature depth profile near peak heating, which is 50 ps after the arrival of the 3 ps pulse. Although we do not know the values of the interface and bulk antiferromagnetic spin Seebeck coefficients, the thermal profile suggests that the interface AF LSSE is more strongly excited than the bulk.

\section{Finite-element simulations of current flow in a cross}

\begin{figure}[!htb]
\centering
\includegraphics[scale=0.7]{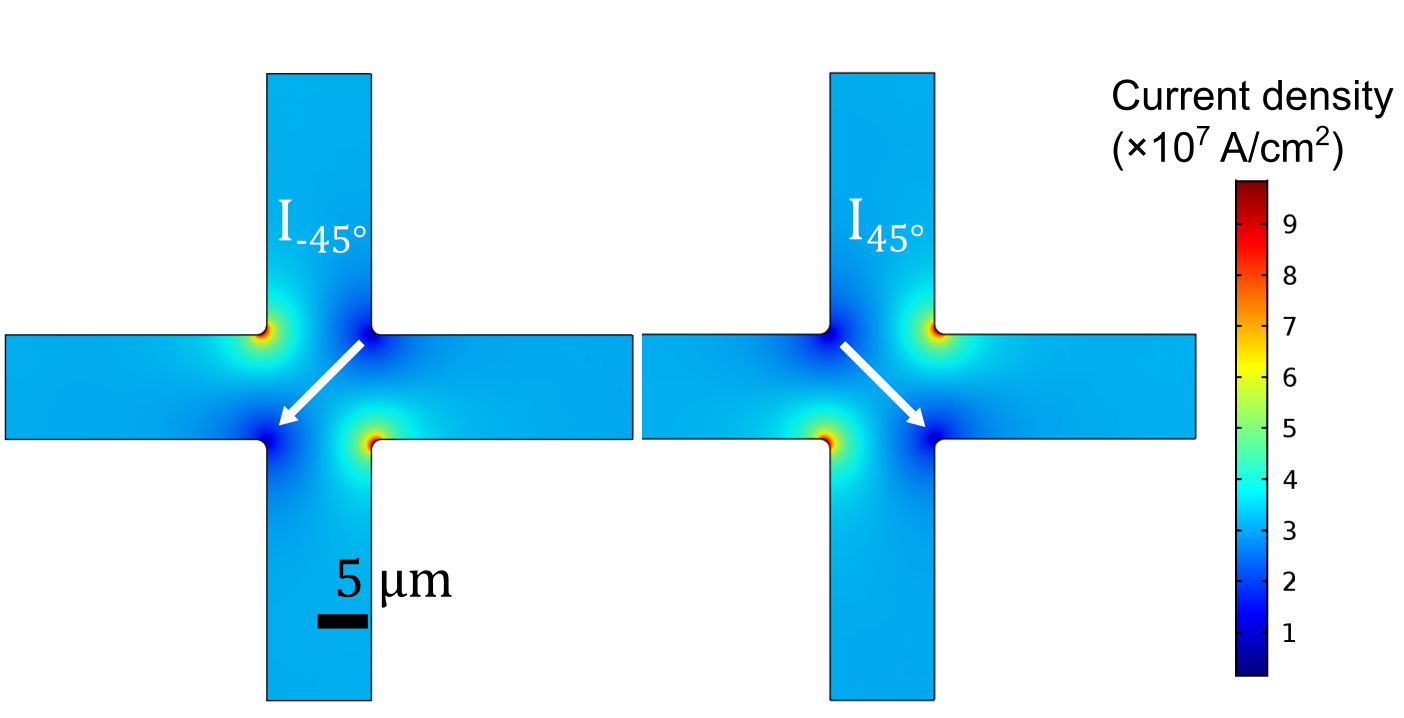}
\caption{Simulated magnitude of current density during the switching process, when the current is applied along a 45$^\circ$ diagonal. We estimate densities of $3.8 \times 10^7 \mathrm{A}/\mathrm{cm}^2$ within the channel, $3.1 \times 10^7 \mathrm{A}/\mathrm{cm}^2$ at the center of the cross, and $8.0 \times 10^7 \mathrm{A}/\mathrm{cm}^2$ at the corner. This is consistent with switching occurring mostly at the sample corner in annealed samples, and also with switching occurring both within the cross channel and at the center in unannealed samples.}
\label{current simulations}
\end{figure}

During the switching process, we apply current to adjacent arms of the cross such that the current flows along a 45$^\circ$ diagonal in the center, which means that the current density is spatially nonuniform. We simulate the spatial current profile in COMSOL in Fig.~\ref{current simulations}. 

We estimate current densities of $3.8 \times 10^7 \mathrm{A}/\mathrm{cm}^2$ within the channel, $3.1 \times 10^7 \mathrm{A}/\mathrm{cm}^2$ at the center of the cross, and $8 \times 10^7 \mathrm{A}/\mathrm{cm}^2$ at the corner. Higher current density at the corner is consistent with the switching profile in annealed NiO, shown in Figure 3 of the main text, where most of the repeatable switching occurs at the corner. Since annealing stabilizes the S-domains in local low-energy configurations, higher current density is necessary to rotate them. Similar current densities within the channel and at the center of the cross are consistent with switching occurring in both regions in unannealed samples. 

\section{Commercial disclaimer}
Certain commercial equipment is identified in this paper to foster understanding. Such identification does not imply recommendation or endorsement by NIST, nor does it imply that the materials or equipment available are necessarily the best available for the purpose.

\bibliographystyle{apsrev4-1}
\bibliography{NiO_with_supplemental_resubmission}% Produces the bibliography via BibTeX. 

\end{document}